\documentclass[twocolumn,superscriptaddress,longbibliography,prc,nofootinbib]{revtex4-2}
\usepackage{times}
\usepackage[T1]{fontenc}
\usepackage{physics}
\usepackage{amsmath,amssymb,mathrsfs}
\usepackage{graphicx}
\usepackage{dcolumn}
\usepackage{bm}
\usepackage{morefloats}
\usepackage{multirow}
\usepackage{mathtools}
\usepackage{xcolor}
\usepackage{longtable}
\usepackage{fix-cm}
\usepackage{verbatim}
\usepackage{float}
\usepackage{siunitx}
\usepackage{kotex}
\usepackage{mathptmx}
\usepackage{subfigure}
\usepackage[colorlinks,allcolors=blue]{hyperref}
\usepackage{adjustbox}

\makeatletter
\def\NAT@def@citea{\def\@citea{\NAT@separator}}
\makeatother

\def\0\\{\nonumber\\}

\newcommand{\beq}{\begin{equation}}
\newcommand{\eeq}{\end{equation}}
\newcommand{\beqn}{\begin{eqnarray}}
\newcommand{\eeqn}{\end{eqnarray}}

\makeatletter
\newcommand\footnoteref[1]{\protected@xdef\@thefnmark{\ref{#1}}\@footnotemark}
\makeatother

\begin{document}


\title{
Impact of hyperon mixing on neutron star structure based on Skyrme-type equations of state:
Systematic analysis of $\Lambda NN$ and $\Lambda\Lambda N$ three-body forces
with Bayesisan inference
}

\author{Taeho Lee}
\email{lee.t.9a69@m.isct.ac.jp}
\affiliation{Department of Physics, School of Science, Institute of Science Tokyo, Tokyo 152-8550, Japan}
\author{Yoonhak Nam}
\affiliation{Department of Physics, School of Science, Institute of Science Tokyo, Tokyo 152-8550, Japan}

\author{Kazuyuki Sekizawa}
\email{sekizawa@phys.sci.isct.ac.jp}
\affiliation{Department of Physics, School of Science, Institute of Science Tokyo, Tokyo 152-8550, Japan}
\affiliation{Nuclear Physics Division, Center for Computational Sciences, University of Tsukuba, Ibaraki 305-8577, Japan}
\affiliation{RIKEN Nishina Center, Saitama 351-0198, Japan}

\date{\today}

\begin{abstract}
\edef\oldrightskip{\the\rightskip}
\begin{description}
\rightskip\oldrightskip\relax
\setlength{\parskip}{0pt} 

\item[Background]
The appearance of $\Lambda$ hyperons in dense neutron-star matter generally softens the equation of state and reduces the maximum mass, while observations of massive neutron stars require sufficient high-density stiffness. Hyperonic interactions must therefore be constrained consistently with both hypernuclear information and neutron-star observables. Density-dependent hyperonic three-body terms provide a possible mechanism for recovering stiffness, but their separate roles in the neutron-star EOS remain to be clarified.

\item[Purpose]
The purpose of this work is to isolate the effects of the effective $\Lambda NN$ and $\Lambda\Lambda N$ density-dependent terms on the hyperonic neutron-star EOS and the resulting mass--radius relation. We aim to clarify how an onset-active term and an onset-inactive but finite-$\Lambda$ term modify the $\Lambda$-onset behavior, the post-onset EOS, the extremum structure of the calculated EOS branch, and the maximum mass.

\item[Methods]
We construct cold beta-equilibrated $npe\mu\Lambda$ matter within a Skyrme energy-density-functional framework and solve the Tolman--Oppenheimer--Volkoff equations for each tabulated EOS. The $\Lambda\Lambda N$ density-dependent term is scanned in the $(\gamma,C_3)$ plane, while the $\Lambda NN$ density-dependent term is scanned separately in the $(\beta,A_3)$ plane. The calculated EOS curves are classified in the $P$-$\varepsilon$ plane according to their monotonicity and extremum structure, and the maps of $N_{\rm ext}^{\rm NS}$ and $N_{\rm ext}^{\rm full}$ are used to interpret the corresponding $M_{\max}$ maps. Representative two-extrema cases are also examined with Maxwell constructions. 
In addition, we perform an exploratory Bayesian analysis using neutron-star mass--radius information and apply XGBoost--SHAP surrogate diagnostics to summarize the parameter sensitivity of selected observables.

\item[Results]
The $\Lambda\Lambda N$ term is onset-inactive in the local continuous onset condition, but it modifies the finite-$\Lambda$ post-onset EOS. Increasing $C_3$ generally enhances the post-onset pressure and raises $M_{\max}$ in mechanically admissible regions, whereas increasing $\gamma$ tends to reduce this enhancement at fixed $C_3$. The comparison of $N_{\rm ext}^{\rm NS}$ and $N_{\rm ext}^{\rm full}$ shows that extrema appearing in the full high-density EOS do not necessarily affect the neutron-star branch relevant to the maximum-mass configuration. By contrast, the $\Lambda NN$ term shifts the $\Lambda$-onset density and modifies the post-onset EOS simultaneously, producing branch-limited regions and Maxwell-construction candidates, especially for the $S\Lambda\Lambda1'$ reference interaction. 
The exploratory Bayesian analysis shows broad posterior degeneracies, but it consistently favors sizable hyperonic three-body repulsion within the adopted mass--radius likelihood. The XGBoost--SHAP analysis identifies the three-body strength parameters, especially $A_3$ and $C_3$, as important controls of $M_{\max}$ and $R_{2.0}$, while $R_{2.0}$ exhibits a broader multi-parameter sensitivity than $M_{\max}$.

\item[Conclusions]
Mass recovery in hyperonic neutron stars is therefore not a single mechanism. In monotonic regions, an increase of $M_{\max}$ can be interpreted as smooth high-density stiffening, whereas in branch-limited or Maxwell-candidate regions the same mass contour can reflect the onset point, the branch limit, or the candidate coexistence structure. The maximum-mass maps must therefore be interpreted together with the onset behavior and the extremum-count maps. Future extensions should include systematic coexistence constructions, including possible Gibbs mixed phases, and simultaneous variations of the $\Lambda NN$ and $\Lambda\Lambda N$ sectors combined with hypernuclear and neutron-star constraints.

\end{description}
\end{abstract}

\maketitle

\section{Introduction}

The equation of state (EOS) of cold dense matter is one of the
central uncertainties in neutron-star physics.  Neutron stars probe
baryonic matter at densities that are not directly accessible in
terrestrial experiments, while their macroscopic properties are
sensitive to the pressure of matter at supra-nuclear density.  The
possible appearance of strange baryons is particularly important in
this context, because hyperon--nucleon and hyperon--hyperon
interactions connect finite hypernuclei with the composition and
stiffness of neutron-star interiors
\cite{gal2016strangeness,tolos2020strangeness}.

Observationally, the existence of neutron stars with masses close to
or above $2M_\odot$ imposes a strong lower bound on the stiffness of
the high-density EOS.  The accurately measured masses of
PSR J1614--2230 and PSR J0348+0432, together with the high-mass
pulsar PSR J0740+6620, provide representative benchmarks for this
constraint \cite{demorest2010j1614,antoniadis2013j0348,fonseca2021j0740}.
Radius information from NICER observations and tidal-deformability
constraints from GW170817 further restrict the pressure of neutron-star
matter at intermediate and high densities
\cite{riley2019j0030,riley2021j0740,choudhury2024j0437,abbott2018gw170817}.
Any EOS including hyperons must therefore be interpreted under the
combined constraints from hypernuclear data and neutron-star
observations.

Hyperons are expected to become energetically favorable in sufficiently
dense beta-equilibrated matter once the relevant chemical-potential
balance is satisfied.  Their appearance generally opens additional
baryonic degrees of freedom and softens the EOS, thereby reducing the
maximum gravitational mass of neutron stars.  This tension between
hyperon formation and the existence of massive neutron stars is usually
referred to as the hyperon puzzle
\cite{balberg1997strangeness,chatterjee2016hyperons}.  A natural way to
reduce this tension is to introduce additional repulsion in the
hyperonic sector at high density, but such repulsion must remain
consistent with empirical information from hypernuclei.

Hypernuclear data provide the primary terrestrial constraints on
hyperonic interactions.  Single-$\Lambda$ binding energies,
$\Lambda$ single-particle spectra, and double-$\Lambda$
hypernuclear information constrain the low- and near-saturation-density
behavior of the $N\Lambda$ and $\Lambda\Lambda$ sectors
\cite{schulze2014,HPL,SLL,SLLd}.  However, direct hyperon--nucleon and
hyperon--hyperon scattering data are scarce, and finite hypernuclei
probe a density regime that is far below the central densities of
massive neutron stars.  Consequently, extrapolations of hyperonic
interactions to neutron-star densities remain model dependent, and it
is important to identify which terms in the EDF control the
$\Lambda$-onset condition, the post-onset EOS, and the resulting
neutron-star observables.

Recent studies have begun to address this problem by combining
hypernuclear information with neutron-star constraints.  Repulsive
$\Lambda NN$ contributions motivated by chiral effective field theory
have been tested against $\Lambda$ binding energies in hypernuclei
and their implications for dense neutron-star matter
\cite{jinno2023repulsiveLambda}.  Bayesian analyses have also been
applied to hyperonic neutron-star matter in relativistic mean-field
models \cite{malik2022hyperonBayes,huang2025hyperonicRMF}.  More
directly related to the present work, recent Skyrme-type studies have
constrained $\Lambda\Lambda$ and $\Lambda\Lambda N$ interactions
using combinations of hypernuclear and astrophysical information
\cite{sun2026hyperonSkyrmeBayes,tanimura2026lambdalambda}.  These studies
demonstrate the importance of combined terrestrial and astrophysical
constraints.  At the same time, a global parameter constraint does not
by itself isolate how each density-dependent hyperonic three-body term
modifies the onset condition, the post-onset EOS branch, and the
maximum mass.

In this work, we employ a Skyrme-type energy-density-functional
framework for hyperonic neutron-star matter.  Skyrme EDFs are useful
for the present purpose because their zero-range and
density-dependent structure allows one to treat finite nuclei,
hypernuclei, and uniform matter within a common phenomenological
framework
\cite{skyrme1959effective,vautherin1972skyrme,rayet1976Hypernuclei,rayet1981LambdaN,schulze2014}.
Here the Skyrme EDF is used as a controlled response framework for
separating the effects of selected hyperonic density-dependent terms,
rather than as a claim of uniqueness for the high-density extrapolation.

The central distinction examined in this work is whether a
density-dependent hyperonic term contributes already at the
$\Lambda$-onset point.  The $\Lambda NN$ density-dependent term is
onset-active, because it modifies the $\Lambda$ chemical potential in
the $n_\Lambda\to0$ limit.  In contrast, the $\Lambda\Lambda N$
density-dependent term considered here is onset-inactive in the local
continuous onset condition, but it can still modify the finite-$\Lambda$
post-onset EOS and the maximum mass.  This distinction concerns only
the onset condition; it does not imply that the $\Lambda\Lambda N$
term is irrelevant for neutron-star structure after a finite
$\Lambda$ fraction develops.

We therefore perform deterministic scans of the hyperonic
density-dependent terms.  The $\Lambda NN$ term is varied in the
$(\beta,A_3)$ plane, whereas the $\Lambda\Lambda N$ term is varied
in the $(\gamma,C_3)$ plane.  For each parameter point, we construct
a beta-equilibrated EOS, solve the Tolman--Oppenheimer--Volkoff (TOV) equations, and evaluate the maximum gravitational mass $M_{\max}$.
In addition, we classify the monotonicity of the calculated
$P(\varepsilon)$ sequence by using the number of extrema in the full EOS and in the neutron-star-relevant pressure range.  This allows us
to distinguish smooth mass recovery from cases in which the maximum mass is affected by branch limitation or by a possible Maxwell-construction candidate.

When a calculated EOS branch becomes nonmonotonic, it cannot be used
as an ordinary single-valued TOV input without further prescription.
We therefore treat one-extremum cases as branch-limited EOSs and
identify two-extrema cases as Maxwell-construction candidates.  The
Maxwell construction is used as a diagnostic for representative cases,
rather than as a replacement for all maximum-mass maps.  In this way,
the deterministic scan separates the smooth stiffening effect of
hyperonic density-dependent repulsion from the effects associated with
nonmonotonic EOS branches.

The deterministic scans in this work are designed to provide a controlled physical baseline rather than to determine a unique optimized hyperonic EDF. They clarify how the $\Lambda NN$ and $\Lambda\Lambda N$ density-dependent terms separately modify the $\Lambda$-onset condition, the post-onset EOS, the extremum structure of $P(\varepsilon)$, and the neutron-star maximum mass. This baseline is then used to interpret the exploratory Bayesian and machine-learning analyses, where neutron-star mass--radius information is used to examine broad parameter tendencies and the sensitivity of selected observables to the hyperonic EDF parameters.

This article is organized as follows. In Sec.~\ref{sec:skyrme_edf_hyperonic_ns}, we introduce the Skyrme-type EDF for hyperonic neutron-star matter and specify the nucleonic, $N\Lambda$, and $\Lambda\Lambda$ sectors used in this work. In Sec.~\ref{sec:ns_structure_eos}, we describe the construction of the beta-equilibrated EOS, the TOV calculation, and the treatment of monotonic and nonmonotonic EOS branches, including the Maxwell-construction diagnostic. In Sec.~\ref{sec:deterministic_effect}, we present deterministic parameter scans for the $\Lambda\Lambda N$ and $\Lambda NN$ terms and discuss their effects on the EOS, branch structure, and maximum mass. In Sec.~\ref{sec:bayesian_ml}, we present the exploratory Bayesian posterior distributions, posterior mass--radius relations, and XGBoost--SHAP sensitivity analysis. Finally, Sec.~\ref{sec:summary_outlook} summarizes the main conclusions.

\section{Skyrme EDF for hyperonic neutron-star matter}
\label{sec:skyrme_edf_hyperonic_ns}

\subsection{Skyrme-type effective interaction and nucleonic baseline}
\label{subsec:skyrme-type_eff_interaction}

Skyrme-type effective interactions provide a compact non-relativistic framework for describing nuclear systems within the Hartree--Fock or energy-density-functional approach \cite{skyrme1959effective,vautherin1972skyrme,bender2003scmf}. 
Their practical advantage is that a small number of zero-range, momentum-dependent, and density-dependent terms can generate an EDF flexible enough to describe nuclear saturation, effective-mass effects, and the density dependence of the EOS. 
In the present work, we consider cold, uniform, spin-saturated matter in neutron-star cores. 
The Skyrme framework is therefore used as the nucleonic baseline on top of which the hyperonic contributions are added. 
The purpose of this subsection is not to review the full Skyrme formalism, but to identify the physical role of each standard $NN$ Skyrme term before introducing the corresponding hyperonic sectors.

The standard $NN$ Skyrme-type effective interaction is written as
\begin{align}
v_{NN}(\mathbf r)
=&\
t_0(1+x_0P_\sigma)\delta(\mathbf r)
\nonumber\\
&+
\frac{1}{2}t_1(1+x_1P_\sigma)
\left[
\mathbf k'^2\delta(\mathbf r)
+
\delta(\mathbf r)\mathbf k^2
\right]
\nonumber\\
&+
t_2(1+x_2P_\sigma)
\mathbf k'\cdot\delta(\mathbf r)\mathbf k
\nonumber\\
&+
\frac{1}{6}t_3(1+x_3P_\sigma)
n_N^\alpha(\mathbf R)\delta(\mathbf r)
\nonumber\\
&+
iW_0
(\boldsymbol{\sigma}_1+\boldsymbol{\sigma}_2)
\cdot
\mathbf k'\times\delta(\mathbf r)\mathbf k .
\label{eq:NN_Skyrme_force}
\end{align}
Equation~\eqref{eq:NN_Skyrme_force} is introduced here to classify the physical role of the five standard Skyrme terms. 
The $t_0$ term is the local zero-range central term. 
In the EDF, it gives the leading momentum-independent density-density contribution and represents the basic central attraction or repulsion in nucleonic matter.

The $t_1$ and $t_2$ terms are momentum-dependent terms. 
After the Hartree--Fock reduction, they generate kinetic-density contributions such as $n_q\tau_q$ and $n_N\tau_N$. 
These terms are associated with the effective-mass structure of the EDF and determine how the kinetic response of dense matter differs from that of a free Fermi gas. 
They are therefore important for the density dependence of the nucleon chemical potentials.

The $t_3$ term is the density-dependent contact term. 
It is used phenomenologically to represent medium and many-body effects that are not captured by the two-body zero-range terms alone. 
This term is essential for nuclear saturation and for controlling the high-density behavior of the nucleonic EOS.

The $W_0$ term is the spin-orbit term. 
It is crucial for shell structure in finite nuclei, but it does not contribute directly to the EOS considered in this work. 
It is nevertheless kept in the classification below because it is part of the standard Skyrme interaction and clarifies which pieces of the full finite-nucleus functional are absent from the neutron-star matter calculation.

For later comparison with the hyperonic sectors, the physical role of each standard $NN$ Skyrme term is summarized in Table~\ref{tab:NN_Skyrme_terms}.
The table follows the structure of the NN block in Ref.~\cite{sun2026hyperonSkyrmeBayes}, where the parameters are classified by channel, type, EDF contribution, and physical meaning.

\begin{table*}[htbp]
\centering
\caption{Classification of the standard $NN$ Skyrme terms used to define the nucleonic baseline. The entries summarize the channel, operator type, leading EDF contribution, and physical role of each term, following the notation of Ref.~\cite{sun2026hyperonSkyrmeBayes}.}
\label{tab:NN_Skyrme_terms}
\begin{adjustbox}{max width=\textwidth}
\begin{tabular}{c c c c c}
\hline\hline
Parameter
& Channel
& Type
& Contribution
& Physical meaning
\\
\hline
$t_0,x_0$
& $NN$
& local, momentum-independent
& $\rho^2$
& central density term
\\
$t_1,x_1$
& $NN$
& non-local, momentum-dependent
& $\rho\tau,\ j^2$
& effective mass / kinetic correlation
\\
$t_2,x_2$
& $NN$
& non-local, momentum-dependent
& $\rho\tau,\ j^2$
& effective mass / kinetic correlation
\\
$t_3,x_3,\alpha$
& $NN$
& density-dependent
& $\rho^{\alpha+2}$
& saturation / medium effect
\\
$W_0$
& $NN$
& spin-orbit
& $\rho\nabla\cdot\mathbf J$
& shell structure in finite nuclei
\\
\hline\hline
\end{tabular}
\end{adjustbox}
\end{table*}

For the neutron-star matter considered here, the spin-orbit and gradient-related terms do not enter the EOS. 
The working nucleonic EDF is therefore obtained from the kinetic term and from the local central, momentum-dependent, and density-dependent contributions that remain in uniform spin-saturated matter. 
In terms of the neutron and proton densities, it can be written as
\begin{align}
\varepsilon_N
=&\
\sum_{q=n,p}
\frac{\hbar^2}{2m_q}\tau_q
\nonumber\\
&+
\frac{t_0}{2}
\left[
\left(1+\frac{x_0}{2}\right)n_N^2
-
\left(\frac{1}{2}+x_0\right)
\left(n_n^2+n_p^2\right)
\right]
\nonumber\\
&+
\frac{1}{4}
\left[
t_1\left(1+\frac{x_1}{2}\right)
+
t_2\left(1+\frac{x_2}{2}\right)
\right]
n_N(\tau_n+\tau_p)
\nonumber\\
&+
\frac{1}{4}
\left[
-t_1\left(\frac{1}{2}+x_1\right)
+
t_2\left(\frac{1}{2}+x_2\right)
\right]
\left(n_n\tau_n+n_p\tau_p\right)
\nonumber\\
&+
\frac{t_3}{12}
n_N^\alpha
\left[
\left(1+\frac{x_3}{2}\right)n_N^2
-
\left(\frac{1}{2}+x_3\right)
\left(n_n^2+n_p^2\right)
\right] .
\label{eq:nucleonic_EDF}
\end{align}
Here, $n_N=n_n+n_p$, and $\tau_q$ is the kinetic density of each nucleon species. 
The first line is the free kinetic contribution. 
The $t_0$ term gives the leading local density-density part, the $t_1$ and $t_2$ terms generate the kinetic-density and effective-mass contributions, and the $t_3$ term gives the density-dependent medium contribution. 
Thus Eq.~\eqref{eq:nucleonic_EDF} is the nucleonic EDF used as the no-$\Lambda$ reference EOS.

This nucleonic baseline fixes the background against which hyperonic
softening and hyperonic density-dependent effects are evaluated. In the
numerical calculations, SLy230b and BSk24 are used as the nucleonic
Skyrme EDFs. These two choices define the nucleonic baselines for the
sector-by-sector hyperonic scans. In this work, the dependence on the
nucleonic EDF is treated as a baseline systematic rather than as a
criterion for ranking different $NN$ Skyrme parametrizations. The same
term-based classification will be used in the next subsection to
organize the $N\Lambda$ and $\Lambda\Lambda$ sectors.

\subsection{\texorpdfstring{Hyperonic Skyrme sectors: $N\Lambda$ and $\Lambda\Lambda$}{Hyperonic Skyrme sectors: N-Lambda and Lambda-Lambda}}
\label{sec:hyperonic_skyrme_sectors}

We now extend the Skyrme-type classification introduced in Sec.~\ref{subsec:skyrme-type_eff_interaction} to the hyperonic part of the EDF.
The hyperonic interaction is organized into the $N\Lambda$ sector and the $\Lambda\Lambda$ sector.
In each sector, the two-body Skyrme-type terms may be supplemented by a density-dependent term, which is used as an effective representation of medium and many-body effects at the EDF level.

The $N\Lambda$ sector describes the interaction between nucleons and a
$\Lambda$ hyperon. Because the $\Lambda$ is an isoscalar baryon and is
distinguishable from nucleons, the $N\Lambda$ channel has no direct
analogue of the isospin-exchange structure appearing in the $NN$
channel. Spin-exchange dependence may still be included
phenomenologically, while possible microscopic channel dependence is
absorbed into the fitted effective $N\Lambda$ parameters. A convenient
Skyrme-type form of the $N\Lambda$ interaction is written as
\cite{rayet1981LambdaN,HPL,sun2026hyperonSkyrmeBayes}
\begin{align}
v_{N\Lambda}(\mathbf r)
=&\
u_0(1+y_0P_\sigma)\delta(\mathbf r)
\nonumber\\
&+
\frac{1}{2}u_1
\left[
\mathbf k'^2\delta(\mathbf r)
+
\delta(\mathbf r)\mathbf k^2
\right]
\nonumber\\
&+
u_2\mathbf k'\cdot\delta(\mathbf r)\mathbf k
\nonumber\\
&+
\frac{3}{8}u_3(1+y_3P_\sigma)
n_N^\beta(\mathbf R)\delta(\mathbf r).
\label{eq:NLambda_force}
\end{align}
The $u_0$ term gives the local central $N\Lambda$ interaction.
The $u_1$ and $u_2$ terms are momentum-dependent terms and generate kinetic-density contributions in the EDF.
The last term is the density-dependent term in the $N\Lambda$ sector, corresponding to an effective $\Lambda NN$ contribution.

After the Hartree--Fock reduction, the $N\Lambda$ contribution to the EDF used here is written as
\begin{align}
\varepsilon_{N\Lambda}
=&\
A_0 n_\Lambda n_N
+
A_1
\left(
n_\Lambda\tau_N+n_N\tau_\Lambda
\right)
+
A_3 n_\Lambda n_N^{1+\beta}.
\label{eq:NLambda_EDF}
\end{align}
Here, $n_N=n_n+n_p$ and $\tau_N=\tau_n+\tau_p$.
The coefficient $A_0$ represents the local central $N\Lambda$ contribution.
The coefficient $A_1$ represents the kinetic-density contribution generated by the momentum-dependent part of the $N\Lambda$ interaction.
The coefficient $A_3$ is the EDF-level strength of the density-dependent term in the $N\Lambda$ sector.
The force-level parameters may be recast into EDF-level coefficients as \cite{schulze2014,schulze2019,sun2026hyperonSkyrmeBayes}
\begin{align}
A_0
&=
u_0\left(1+\frac{y_0}{2}\right),
\label{eq:A0_relation}
\\
A_1
&=
\frac{u_1+u_2}{4},
\label{eq:A1_relation}
\\
A_3
&=
\frac{3}{8}u_3\left(1+\frac{y_3}{2}\right).
\label{eq:A3_relation}
\end{align}
The finite-nucleus gradient coefficient, usually denoted by $A_2$, is not included in Eq.~\eqref{eq:NLambda_EDF}.
It is nevertheless listed below when an original parameter set is quoted in the EDF-coefficient convention.
In the parameter scans below, the pair $(\beta,A_3)$ specifies the density-dependent term in the $N\Lambda$ sector.

The $\Lambda\Lambda$ sector describes the interaction between two $\Lambda$ hyperons.
Since the $\Lambda$ hyperon is isoscalar, no isospin-exchange structure analogous to the $NN$ channel is introduced in the $\Lambda\Lambda$ interaction.
The standard Skyrme-type form of the $\Lambda\Lambda$ interaction is written as
\cite{SLL,SLLd,sun2026hyperonSkyrmeBayes}.
\begin{align}
v_{\Lambda\Lambda}(\mathbf r)
=&\
\lambda_0\delta(\mathbf r)
\nonumber\\
&+
\frac{1}{2}\lambda_1
\left[
\mathbf k'^2\delta(\mathbf r)
+
\delta(\mathbf r)\mathbf k^2
\right]
\nonumber\\
&+
\lambda_2\mathbf k'\cdot\delta(\mathbf r)\mathbf k
\nonumber\\
&+
\lambda_3 n_N^\gamma(\mathbf R)\delta(\mathbf r).
\label{eq:LLambda_force}
\end{align}
The $\lambda_0$ term gives the local central $\Lambda\Lambda$ interaction.
The $\lambda_1$ and $\lambda_2$ terms are momentum-dependent terms.
The last term is the density-dependent term in the $\Lambda\Lambda$ sector, corresponding to an effective $\Lambda\Lambda N$ contribution.

The corresponding $\Lambda\Lambda$ contribution to the EDF is written as
\begin{align}
\varepsilon_{\Lambda\Lambda}
=&\
C_0 n_\Lambda^2
+
C_1 n_\Lambda\tau_\Lambda
+
C_3 n_\Lambda^2 n_N^\gamma .
\label{eq:LLambda_EDF}
\end{align}
The coefficient $C_0$ represents the local central $\Lambda\Lambda$ contribution.
The coefficient $C_1$ represents the kinetic-density contribution generated by the momentum-dependent part of the $\Lambda\Lambda$ interaction.
The coefficient $C_3$ is the EDF-level strength of the density-dependent term in the $\Lambda\Lambda$ sector.
For the general force in Eq.~\eqref{eq:LLambda_force}, the corresponding coefficients are
\begin{align}
C_0
&=
\frac{\lambda_0}{4},
\label{eq:C0_relation}
\\
C_1
&=
\frac{\lambda_1+3\lambda_2}{8},
\label{eq:C1_relation}
\\
C_3
&=
\frac{\lambda_3}{4}.
\label{eq:C3_relation}
\end{align}

For the S$\Lambda\Lambda$ parameter sets adopted as the two-body $\Lambda\Lambda$ baseline in this work, the original fits set $\lambda_2=\lambda_3=0$ \cite{SLL}.
This choice is motivated by the fact that p-wave contributions do not participate in the lowest $\Lambda$ single-particle level, while the $\Lambda\Lambda-\Xi N$ coupling associated with the density-dependent term is expected to be strongly suppressed by Pauli blocking in heavy nuclei \cite{SLLd}.
With this baseline convention, $C_1=\lambda_1/8$ and $C_3=0$ for the fitted two-body $\Lambda\Lambda$ force.
In the present neutron-star calculation, however, this convention is used only to define the baseline $\Lambda\Lambda$ parameter sets.
The density-dependent coefficient $C_3$ is then introduced as an additional EDF-level parameter and varied together with $\gamma$ to test the sensitivity of the EOS to an effective $\Lambda\Lambda N$ contribution.
Thus, $\lambda_2=\lambda_3=0$ should be understood as a fit-level truncation of the baseline $\Lambda\Lambda$ force, not as a statement that such contributions are absent in dense matter.

The hyperonic EDF terms introduced above are summarized in Table~\ref{tab:hyperonic_terms_classification}.
Following the classification scheme in Ref.~\cite{sun2026hyperonSkyrmeBayes}, the entries are organized by parameter, channel, type, contribution, and physical meaning.

\begin{table*}[htbp]
\centering
\caption{Classification of the hyperonic Skyrme terms in the $N\Lambda$ and $\Lambda\Lambda$ sectors. The density-dependent $N\Lambda$ term corresponds to an effective $\Lambda NN$ contribution, whereas the density-dependent $\Lambda\Lambda$ term corresponds to an effective $\Lambda\Lambda N$ contribution at the EDF level.}
\label{tab:hyperonic_terms_classification}
\begin{adjustbox}{max width=\textwidth}
\begin{tabular}{c c c c c}
\hline\hline
Parameter
& Channel
& Type
& Contribution
& Physical meaning
\\
\hline
$u_0,y_0$
& $N\Lambda$
& Local, momentum-independent
& $\rho_N\rho_\Lambda$
& Central attraction between $N$ and $\Lambda$
\\
$u_1,u_2$
& $N\Lambda$
& Non-local, momentum-dependent
& $\rho_N\tau_\Lambda+\rho_\Lambda\tau_N$
& Effective mass of $\Lambda$, kinetic feedback
\\
$y_1,y_2$
& $N\Lambda$
& Non-local, momentum-dependent
& Exchange terms
& Spin/isospin dependence
\\
$u_3,y_3,\beta$
& $N\Lambda$
& Density-dependent
& $\rho_N^\beta\rho_\Lambda$
& Medium-induced repulsion
\\
$\lambda_0$
& $\Lambda\Lambda$
& Local, momentum-independent
& $\rho_\Lambda^2$
& Central attraction/repulsion in pure $\Lambda$ matter
\\
$\lambda_1,\lambda_2$
& $\Lambda\Lambda$
& Non-local, momentum-dependent
& $\rho_\Lambda\tau_\Lambda,\ j^2$
& $\Lambda$ effective mass, dynamical corrections
\\
$\lambda_3,\gamma$
& $\Lambda\Lambda$
& Density-dependent
& $\rho_\Lambda^2\rho_N^\gamma$
& Repulsive many-body effects at high density
\\
\hline\hline
\end{tabular}
\end{adjustbox}
\end{table*}

The classification in Table~\ref{tab:hyperonic_terms_classification} specifies the physical role of each Skyrme term, whereas the numerical calculations require a choice of hyperonic parameter sets.
The $N\Lambda$ parameter sets used in this work are summarized in Table~\ref{tab:NLambda_parameter_sets}.
The $\Lambda\Lambda$ two-body parameter sets used as the baseline of the $\Lambda\Lambda$ sector are summarized in Table~\ref{tab:LLambda_parameter_sets}.

\begin{table*}[htbp]
\centering
\caption{$N\Lambda$ Skyrme parameter sets used in this work. 
The upper block lists force-level parameters, whereas the lower block lists the SLL4 and SLL4$^{\prime}$ sets in the EDF-coefficient convention. 
The dimensions are shown in the column headings for dimensional quantities; the density exponent $\beta$ and the spin-exchange parameters $y_0$ and $y_3$ are dimensionless. 
The coefficient $A_2$ is retained only for identifying the original finite-nucleus parametrization and does not enter the uniform-matter EOS used in this work.}
\label{tab:NLambda_parameter_sets}
\begin{adjustbox}{max width=\textwidth}
\begin{tabular*}{\textwidth}{@{\extracolsep{\fill}} c c c c c c c c @{}}
\hline\hline
$N\Lambda$ models 
& $\beta$
& \begin{tabular}{@{}c@{}} $u_0$ \\ $[\mathrm{MeV\,fm}^{3}]$ \end{tabular}
& \begin{tabular}{@{}c@{}} $u_1$ \\ $[\mathrm{MeV\,fm}^{5}]$ \end{tabular}
& \begin{tabular}{@{}c@{}} $u_2$ \\ $[\mathrm{MeV\,fm}^{5}]$ \end{tabular}
& \begin{tabular}{@{}c@{}} $u_3$ \\ $[\mathrm{MeV\,fm}^{3\beta+3}]$ \end{tabular}
& $y_0$
& $y_3$
\\
\hline
YBZ5 \cite{YBZ}
& 1 
& $-$315.3 
& 23.14 
& $-$23.14 
& 2000.0 
& $-$0.109 
& 0 
\\
YBZ6 \cite{YBZ}
& 1 
& $-$372.2 
& 100.4 
& 79.60 
& 2000.0 
& $-$0.107 
& 0 
\\
HP$\Lambda$2 \cite{HPL}
& 1 
& $-$399.946 
& 83.426 
& 11.455 
& 2046.818 
& $-$0.486 
& $-$0.660 
\\
O$\Lambda$1 \cite{HPL}
& 1 
& $-$236.5835 
& 116.8704 
& $-$112.8812 
& 1453.3493 
& 0.1271 
& $-$0.3110 
\\
N$\Lambda$1 \cite{HPL}
& 1 
& $-$253.3250 
& 147.1264 
& $-$83.5843 
& 1684.9876 
& 0.5802 
& 0.4831 
\\
N$\Lambda$2 \cite{HPL}
& $1/3$ 
& $-$518.620 
& 82.0944 
& $-$19.9772 
& 1102.2221 
& $-$0.3854 
& $-$0.5645 
\\
\hline
$N\Lambda$ models 
& $\beta$
& \begin{tabular}{@{}c@{}} $A_0$ \\ $[\mathrm{MeV\,fm}^{3}]$ \end{tabular}
& \begin{tabular}{@{}c@{}} $A_1$ \\ $[\mathrm{MeV\,fm}^{5}]$ \end{tabular}
& \begin{tabular}{@{}c@{}} $A_2$ \\ $[\mathrm{MeV\,fm}^{5}]$ \end{tabular}
& \begin{tabular}{@{}c@{}} $A_3$ \\ $[\mathrm{MeV\,fm}^{3\beta+3}]$ \end{tabular}
& \multicolumn{2}{c}{}
\\
\hline
SLL4\cite{schulze2014}
& 1 
& $-$316.0 
& 23.25 
& 13.88 
& 650.0 
& \multicolumn{2}{c}{}
\\
SLL4$^{\prime}$\cite{schulze2014}
& 1 
& $-$326.0 
& 20.50 
& 20.75 
& 705.0 
& \multicolumn{2}{c}{}
\\
\hline\hline
\end{tabular*}
\end{adjustbox}
\end{table*}

Although $A_2$ is listed for the SLL4 and SLL4$^{\prime}$ sets because these interactions are originally specified in an EDF-coefficient convention, it does not enter Eq.~\eqref{eq:NLambda_EDF}.
The coefficient is therefore retained only for identifying the original parameter set.

\begin{table}[htbp]
\centering
\caption{$\Lambda\Lambda$ Skyrme parameter sets used as baseline two-body $\Lambda\Lambda$ interactions. 
All listed fitted sets have $\lambda_2=\lambda_3=0$, so the density-dependent exponent $\gamma$ is not active in the baseline force. }
\label{tab:LLambda_parameter_sets}
\begin{tabular}{c c c c c c}
\hline\hline
\multirow{2}{*}{$\Lambda\Lambda$ set}
& \multirow{2}{*}{$\gamma$}
& $\lambda_0$
& $\lambda_1$
& $\lambda_2$
& $\lambda_3$
\\
&
&
$[\mathrm{MeV\,fm}^{3}]$
& $[\mathrm{MeV\,fm}^{5}]$
& $[\mathrm{MeV\,fm}^{5}]$
& $[\mathrm{MeV\,fm}^{3\gamma+3}]$
\\
\hline
S$\Lambda\Lambda$1 \cite{SLL}
& -- 
& $-$312.6 
& 57.5 
& 0
& 0
\\
S$\Lambda\Lambda$2 \cite{SLL}
& -- 
& $-$437.7 
& 240.7 
& 0
& 0
\\
S$\Lambda\Lambda$3 \cite{SLL}
& -- 
& $-$831.8 
& 922.9 
& 0
& 0
\\
S$\Lambda\Lambda$1$^{\prime}$ \cite{SLLd}
& -- 
& $-$37.9 
& 14.1 
& 0
& 0
\\
S$\Lambda\Lambda$3$^{\prime}$ \cite{SLLd}
& -- 
& $-$156.4 
& 347.2 
& 0
& 0
\\
\hline\hline
\end{tabular}
\end{table}

Tables~\ref{tab:NLambda_parameter_sets} and \ref{tab:LLambda_parameter_sets} specify the hyperonic parameter sets used to construct the baseline EOSs. In the discussion below, these sets are not used with equal emphasis. For the $N\Lambda$ sector, we mainly use the SLL4-type interactions, because they were constructed to reproduce single-$\Lambda$ hypernuclear binding information over a broad mass range from light to heavy hypernuclei \cite{schulze2014}. Among them, SLL4$^{\prime}$ is used as the primary $N\Lambda$ reference, while SLL4 is retained as a closely related comparison set. For the $\Lambda\Lambda$ sector, we distinguish the original S$\Lambda\Lambda1$--S$\Lambda\Lambda3$ sets from the updated prime sets S$\Lambda\Lambda1^{\prime}$ and S$\Lambda\Lambda3^{\prime}$. The prime sets are treated as the main $\Lambda\Lambda$-sector references because they correspond to the weaker bond energy assigned to the $^{13}_{\Lambda\Lambda}\mathrm{B}$ ground state after taking into account the excited daughter single-$\Lambda$ hypernucleus $^{13}_{\Lambda}\mathrm{C}^{*}$ in the decay channel \cite{SLLd}. In the deterministic scans below, the density-dependent coefficients $A_3$ and $C_3$, together with their density exponents $\beta$ and $\gamma$, are varied to examine the effect of hyperonic density-dependent terms on the EOS and neutron-star structure. In the next subsection, the interaction contributions introduced here are combined with the nucleonic and leptonic parts to construct the total energy density.

\subsection{Total energy density of hyperonic neutron-star matter}
\label{subsec:total_energy_density_hyperonic_ns_matter}

The total energy density used for the EOS is constructed from the baryon rest-mass contribution, the nucleonic EDF, the hyperonic contribution, and the lepton contribution.
We write
\begin{align}
\varepsilon
=
\sum_{b=n,p,\Lambda}
m_b n_b
+
\varepsilon_N
+
\varepsilon_\Lambda
+
\varepsilon_l .
\label{eq:total_energy_density}
\end{align}
Here, $\varepsilon_N$ is the nucleonic EDF defined in Eq.~\eqref{eq:nucleonic_EDF}.
The first term in Eq.~\eqref{eq:total_energy_density} adds the baryon rest-mass contribution explicitly.

The hyperonic part is defined as
\begin{align}
\varepsilon_\Lambda
=
\frac{\hbar^2}{2m_\Lambda}\tau_\Lambda
+
\varepsilon_{N\Lambda}
+
\varepsilon_{\Lambda\Lambda}.
\label{eq:hyperonic_energy_density}
\end{align}
The first term is the free kinetic-energy contribution of the $\Lambda$ particles.
The terms $\varepsilon_{N\Lambda}$ and $\varepsilon_{\Lambda\Lambda}$ are given in Eqs.~\eqref{eq:NLambda_EDF} and \eqref{eq:LLambda_EDF}, respectively.
The density-dependent term in $\varepsilon_{N\Lambda}$ represents the effective $\Lambda NN$ contribution within the $N\Lambda$ sector, while the density-dependent term in $\varepsilon_{\Lambda\Lambda}$ represents the effective $\Lambda\Lambda N$ contribution within the $\Lambda\Lambda$ sector.

The total baryon number density is
\begin{align}
n_B
=
n_n+n_p+n_\Lambda .
\label{eq:baryon_number_density}
\end{align}
At zero temperature, the kinetic density of each baryon species is fixed by its Fermi momentum and hence by its number density:
\begin{align}
\tau_i
=
\frac{3}{5}
\left(3\pi^2\right)^{2/3}
n_i^{5/3},
\qquad
i=n,p,\Lambda .
\label{eq:kinetic_density}
\end{align}
Thus, once the particle fractions are determined, the baryonic part of the energy density is fixed by the particle densities.

Electrons and muons are included as free relativistic Fermi gases \cite{glendenning2000compactStars,oertel2017eosReview}.
The total lepton energy density is defined as
\begin{align}
\varepsilon_l
=
\sum_{l=e,\mu}
\frac{1}{\pi^2}
\int_0^{k_{F,l}}
k^2
\sqrt{k^2+m_l^2}
\,dk,
\qquad
k_{F,l}
=
(3\pi^2 n_l)^{1/3}.
\label{eq:lepton_energy_density}
\end{align}
Here, $n_l$, $m_l$, and $k_{F,l}$ are the number density, mass, and Fermi momentum of the lepton species $l$, respectively.
If muons are absent at a given density, the corresponding contribution is set to zero by taking $n_\mu=0$.
The lepton rest-mass contributions are already included in Eq.~\eqref{eq:lepton_energy_density} through the relativistic single-particle energy.

In this convention, baryon rest masses are included explicitly in Eq.~\eqref{eq:total_energy_density}, while the lepton rest-mass contributions are already included in Eq.~\eqref{eq:lepton_energy_density}.
The chemical potentials, pressure, charge neutrality, and beta-equilibrium conditions are specified in the next subsection; these equations determine the particle fractions, and hence the composition, of neutron-star matter.

\subsection{Chemical potentials, pressure, and beta equilibrium}
\label{subsec:chem_potential}

The particle fractions are determined by minimizing the total energy density under the constraints appropriate for cold neutron-star matter.
When all species are included, the composition may be described by the six density variables
$(n_B,n_n,n_p,n_\Lambda,n_e,n_\mu)$.
Baryon-number conservation, charge neutrality, and beta-equilibrium conditions provide five relations among these variables.
The remaining independent variable is taken to be $n_B$, and the equilibrium particle fractions $Y_i=n_i/n_B$ are obtained along the resulting one-parameter sequence.

The chemical potential of each species is defined from the total energy density in Eq.~\eqref{eq:total_energy_density} as
\begin{align}
\mu_i
=
\left.
\frac{\partial \varepsilon}{\partial n_i}
\right|_{\{n_{j\neq i}\}},
\qquad
i=n,p,\Lambda,e,\mu .
\label{eq:chemical_potential}
\end{align}
The derivatives in Eq.~\eqref{eq:chemical_potential} include the implicit density dependence of the kinetic densities $\tau_i(n_i)$, as well as the rearrangement contributions from the density-dependent terms in the EDF.
For the leptons, Eq.~\eqref{eq:chemical_potential} gives
\begin{align}
\mu_l
=
\sqrt{k_{F,l}^2+m_l^2},
\qquad
l=e,\mu .
\label{eq:lepton_chemical_potential}
\end{align}

The pressure is obtained from the thermodynamic relation \cite{glendenning2000compactStars,sun2026hyperonSkyrmeBayes}.
\begin{align}
P
=
\sum_{i=n,p,\Lambda,e,\mu}
n_i\mu_i
-
\varepsilon .
\label{eq:pressure}
\end{align}
Here, $\varepsilon$ and $\mu_i$ are both defined with the rest-mass convention specified in Sec.~\ref{subsec:total_energy_density_hyperonic_ns_matter}.
After imposing the equilibrium conditions below, Eq.~\eqref{eq:pressure} gives the EOS as a relation between $P$ and $\varepsilon$.
This $P(\varepsilon)$ relation is used as the input for the neutron-star structure calculation discussed later.

At a given $n_B$, baryon-number conservation is written as
\begin{align}
n_B
=
n_n+n_p+n_\Lambda .
\label{eq:baryon_number_constraint}
\end{align}
Charge neutrality requires
\begin{align}
n_p
=
n_e+n_\mu .
\label{eq:charge_neutrality}
\end{align}
For neutrino-free beta-equilibrated matter, the weak-interaction equilibrium conditions are
\begin{align}
\mu_n
&=
\mu_p+\mu_e,
\nonumber\\
\mu_e
&=
\mu_\mu,
\nonumber\\
\mu_n
&=
\mu_\Lambda .
\label{eq:beta_equilibrium}
\end{align}
The first condition corresponds to neutron--proton conversion, the second to electron--muon equilibrium, and the third to chemical equilibrium between neutrons and $\Lambda$ hyperons.
If muons are not populated at a given density, the solution is taken with $n_\mu=0$; the muon component appears once the electron chemical potential reaches the muon rest mass.

The last condition in Eq.~\eqref{eq:beta_equilibrium} is written as $\mu_n=\mu_\Lambda$ because the chemical potentials in Eq.~\eqref{eq:chemical_potential} are defined from the rest-mass-inclusive energy density.
If one instead uses baryonic chemical potentials with the rest masses subtracted, the same physical condition is written as
$\mu_n+m_n=\mu_\Lambda+m_\Lambda$, as in Ref.~\cite{sun2026hyperonSkyrmeBayes}.
Thus the difference is purely a matter of convention.

The first appearance of $\Lambda$ hyperons is determined by the boundary condition at $n_\Lambda=0$.
Before the onset, the solution is restricted to $n_\Lambda=0$, and the relevant threshold chemical potential is
\begin{align}
\mu_\Lambda^{(0)}
=
\left.
\frac{\partial \varepsilon}{\partial n_\Lambda}
\right|_{n_\Lambda=0}.
\label{eq:lambda_threshold_chemical_potential}
\end{align}
The $\Lambda$ onset occurs when
\begin{align}
\mu_n
=
\mu_\Lambda^{(0)} .
\label{eq:lambda_onset_condition}
\end{align}
Above this threshold, $n_\Lambda$ is determined self-consistently by the full set of equilibrium conditions in Eq.~\eqref{eq:beta_equilibrium}.
The dependence of this onset condition on the $N\Lambda$ and $\Lambda\Lambda$ sectors is discussed in the next subsection.

Equations~\eqref{eq:chemical_potential}--\eqref{eq:beta_equilibrium} determine the equilibrium composition for each $n_B$.
Once the composition is obtained, Eqs.~\eqref{eq:total_energy_density} and \eqref{eq:pressure} define the EOS.
The sector dependence of the $\Lambda$-onset condition and its implication for the EOS are discussed next.
\subsection{\texorpdfstring{$\Lambda$-onset condition and sector dependence}{Lambda-onset condition and sector dependence}}
\label{sec:onset_condition}
The $\Lambda$-onset density is defined as the density at which a solution with a finite $\Lambda$ fraction first becomes energetically allowed.
For a continuous onset at zero temperature, this condition is obtained by evaluating the $\Lambda$ chemical potential at the boundary $n_\Lambda=0$.
Before the onset, the equilibrium solution is restricted to $n_\Lambda=0$, and the relevant chemical potential is
\begin{align}
\mu_\Lambda^{(0)}
=
\left.
\frac{\partial \varepsilon}{\partial n_\Lambda}
\right|_{n_\Lambda=0} .
\label{eq:lambda_chemical_potential_onset}
\end{align}
Here, the derivative is taken with the other particle densities fixed.
With the rest-mass convention adopted in Sec.~\ref{subsec:total_energy_density_hyperonic_ns_matter}, the onset condition is
\begin{align}
\mu_n
=
\mu_\Lambda^{(0)} .
\end{align}
This is the threshold condition for $\Lambda$ appearance, written in the rest-mass-inclusive convention used in this work.

To see which hyperonic sector enters Eq.~\eqref{eq:lambda_chemical_potential_onset}, we rewrite the two hyperonic EDF contributions:
\begin{align}
\varepsilon_{N\Lambda}
=&\
A_0 n_\Lambda n_N
+
A_1
\left(
n_\Lambda\tau_N+n_N\tau_\Lambda
\right)
+
A_3 n_\Lambda n_N^{1+\beta},
\label{eq:NLambda_EDF_onset_rewrite}
\\
\varepsilon_{\Lambda\Lambda}
=&\
C_0 n_\Lambda^2
+
C_1 n_\Lambda\tau_\Lambda
+
C_3 n_\Lambda^2 n_N^\gamma .
\label{eq:LLambda_EDF_onset_rewrite}
\end{align}
The $N\Lambda$ contribution gives
\begin{align}
\frac{\partial \varepsilon_{N\Lambda}}{\partial n_\Lambda}
=
A_0 n_N
+
A_1
\left(
\tau_N
+
n_N
\frac{\partial \tau_\Lambda}{\partial n_\Lambda}
\right)
+
A_3 n_N^{1+\beta}.
\label{eq:dNLambda_dnLambda}
\end{align}
Since $\tau_\Lambda\propto n_\Lambda^{5/3}$, one has
$\partial \tau_\Lambda/\partial n_\Lambda\propto n_\Lambda^{2/3}$, which vanishes at $n_\Lambda=0$.
Therefore,
\begin{align}
\left.
\frac{\partial \varepsilon_{N\Lambda}}{\partial n_\Lambda}
\right|_{n_\Lambda=0}
=
A_0 n_N
+
A_1\tau_N
+
A_3 n_N^{1+\beta}.
\label{eq:dNLambda_dnLambda_onset}
\end{align}
Thus the local central, momentum-dependent, and density-dependent terms in the $N\Lambda$ sector directly enter the $\Lambda$-onset condition.

By contrast, the derivative of the $\Lambda\Lambda$ contribution is
\begin{align}
\frac{\partial \varepsilon_{\Lambda\Lambda}}{\partial n_\Lambda}
=
2C_0 n_\Lambda
+
C_1
\left(
\tau_\Lambda
+
n_\Lambda
\frac{\partial \tau_\Lambda}{\partial n_\Lambda}
\right)
+
2C_3 n_\Lambda n_N^\gamma .
\label{eq:dLLambda_dnLambda}
\end{align}
Every term in Eq.~\eqref{eq:dLLambda_dnLambda} vanishes at $n_\Lambda=0$, and hence
\begin{align}
\left.
\frac{\partial \varepsilon_{\Lambda\Lambda}}{\partial n_\Lambda}
\right|_{n_\Lambda=0}
=
0 .
\label{eq:dLLambda_dnLambda_onset}
\end{align}
The $\Lambda\Lambda$ sector therefore does not directly affect the local continuous $\Lambda$-onset condition.
Equivalently, changing the $\Lambda\Lambda$ interaction or the density-dependent $\Lambda\Lambda N$-type term cannot shift the threshold condition in Eq.~\eqref{eq:lambda_onset_condition} at $n_\Lambda=0$.

Combining these results with the free $\Lambda$ kinetic term, whose derivative also vanishes at $n_\Lambda=0$, gives
\begin{align}
\mu_\Lambda^{(0)}
=
m_\Lambda
+
A_0 n_N
+
A_1\tau_N
+
A_3 n_N^{1+\beta}.
\label{eq:lambda_onset_mu_explicit}
\end{align}
The quantities $n_N$ and $\tau_N$ in Eq.~\eqref{eq:lambda_onset_mu_explicit} are evaluated on the $n_\Lambda=0$ equilibrium branch.
The onset density is then determined by the first density at which Eq.~\eqref{eq:lambda_onset_condition} is satisfied.

This distinction should be interpreted only as a statement about the onset condition.
It does not imply that the $\Lambda\Lambda$ sector is irrelevant for the EOS.
Once a finite $\Lambda$ fraction is present, the terms in $\varepsilon_{\Lambda\Lambda}$ contribute to the chemical potentials and pressure, and therefore affect the high-density behavior of the EOS.
The strict qualitative difference is that the $N\Lambda$ sector is onset-active, whereas the $\Lambda\Lambda$ sector contributes only after the $\Lambda$ fraction becomes finite.
This distinction provides the formal basis for interpreting the sector-dependent EOS behavior discussed in the results.

\section{Neutron-star structure calculation and EOS branches}
\label{sec:ns_structure_eos}

\subsection{TOV equations and EOS input}

The neutron-star structure calculation requires the EOS as a relation between pressure and energy density, $P(\varepsilon)$.
Given this relation, the mass--radius sequence is obtained by solving the Tolman--Oppenheimer--Volkoff equations \cite{tolman1939,oppenheimer1939}.
Throughout this section, we use units $G=c=1$.
The TOV equations are written as
\begin{align}
\frac{\dd P}{\dd r}
&=
-
\frac{
\left[\varepsilon(r)+P(r)\right]
\left[
m(r)+4\pi r^3P(r)
\right]
}{
r\left[r-2m(r)\right]
},
\label{eq:TOV_pressure}
\\
\frac{\dd m}{\dd r}
&=
4\pi r^2\varepsilon(r).
\label{eq:TOV_mass}
\end{align}
Here, $m(r)$ is the gravitational mass enclosed within radius $r$, and $P(r)$ and $\varepsilon(r)$ are the local pressure and energy density.
The integration starts from a central pressure $P_c$ with
\begin{align}
P(0)=P_c,
\qquad
m(0)=0 .
\label{eq:TOV_center_boundary}
\end{align}
The radius $R$ is defined by the surface condition
\begin{align}
P(R)=0,
\label{eq:TOV_surface}
\end{align}
and the corresponding gravitational mass is
\begin{align}
M=m(R).
\label{eq:TOV_mass_radius}
\end{align}
By varying the central pressure $P_c$, one obtains a sequence of neutron-star configurations and hence the mass--radius relation.
The maximum mass $M_{\max}$ is then identified from this sequence.

For direct TOV integration, the EOS must specify the energy density corresponding to the local pressure encountered during the integration.
For an ordinary continuous EOS, this is simply a single-valued $\varepsilon(P)$ relation.
If the calculated equilibrium sequence is nonmonotonic, an additional prescription is required before the EOS can be used in the TOV equations.
This motivates the construction of the EOS sequence and the classification of monotonic and nonmonotonic EOS branches in the following subsections.

\subsection{Construction of the EOS sequence}
\label{subsec:construction_eos_sequence}

The EOS sequence is constructed by solving the equilibrium conditions specified in Sec.~\ref{subsec:chem_potential}.
Before choosing the total baryon density, the composition is described by the six density variables
$(n_B,n_n,n_p,n_\Lambda,n_e,n_\mu)$.
Baryon-number conservation, charge neutrality, and the three beta-equilibrium conditions provide five relations among them.
Thus, the remaining degree of freedom can be chosen as $n_B$, and the equilibrium composition is obtained as a one-parameter sequence labeled by $n_B$.

For each prescribed value of $n_B$, the particle densities
$
(n_n,n_p,n_\Lambda,n_e,n_\mu)
$
are determined by solving the constraints in Eqs.~\eqref{eq:baryon_number_constraint}--\eqref{eq:beta_equilibrium}.
Below the $\Lambda$-onset density, the solution is restricted to $n_\Lambda=0$, and the onset condition is monitored through Eq.~\eqref{eq:lambda_onset_condition}.
Above the onset, $n_\Lambda$ is determined self-consistently together with the other particle densities.

Once the equilibrium composition is obtained, the total energy density and pressure are evaluated from Eqs.~\eqref{eq:total_energy_density} and \eqref{eq:pressure}.
This gives the EOS as a parametric sequence,
\begin{align}
n_B
\longmapsto
\left[
\varepsilon(n_B),P(n_B)
\right].
\label{eq:EOS_parametric_sequence}
\end{align}
The relation $P(\varepsilon)$ required in the TOV equations is then obtained from this equilibrium sequence.

For a monotonic sequence, this construction directly provides the EOS used in the TOV integration.
However, some hyperonic EOSs obtained in the present calculation develop nonmonotonic behavior after the appearance of $\Lambda$ hyperons.
In such cases, the parametric sequence in Eq.~\eqref{eq:EOS_parametric_sequence} must be classified before it can be used as an EOS input.
The next subsection classifies the calculated EOS branches according to their monotonicity and the number of extrema.

\subsection{Monotonic and nonmonotonic EOS branches}
\label{sec:eos_branches}

The EOS sequence constructed in Sec.~\ref{subsec:construction_eos_sequence} is first obtained as a parametric curve,
\begin{align}
n_B \mapsto [\varepsilon(n_B),P(n_B)] .
\end{align}
For the TOV calculation, this sequence must provide an admissible relation between the pressure and the energy density.
In this subsection, an EOS branch denotes a connected segment of the calculated parametric curve in the $P$-$\varepsilon$ plane.
Before using the sequence in the TOV calculation, we classify the branch according to the monotonicity of $P(\varepsilon)$ and the number of extrema.

An extremum is defined as a local maximum or minimum of $P$ as a function of $\varepsilon$, satisfying
\begin{align}
\frac{\dd P}{\dd \varepsilon}
=
0 .
\label{eq:EOS_extremum_condition}
\end{align}
We denote the number of extrema in a calculated EOS sequence by $N_{\rm ext}$.
This classification is used as an operational prescription for treating the calculated EOS sequence before the TOV integration.

\begin{figure}[t]
    \centering
    \includegraphics[width=\columnwidth]{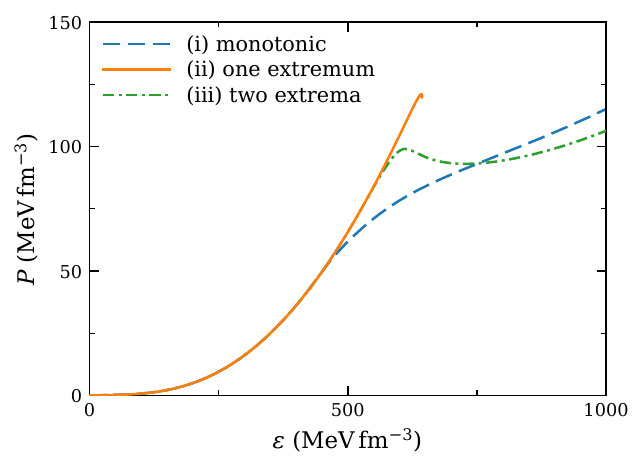}
    \caption{
    Representative EOS branches in the $P$-$\varepsilon$ plane.
    Curve (i) shows a monotonic branch with $N_{\rm ext}=0$.
    Curve (ii) shows a nonmonotonic branch with one extremum, $N_{\rm ext}=1$, for which the EOS is used only up to the first extremum in the TOV calculation.
    Curve (iii) shows a nonmonotonic branch with two extrema, $N_{\rm ext}=2$, where the negative-slope segment is bounded by two monotonic branches and is treated as a Maxwell-construction candidate.
    The displayed examples are baseline hyperonic EOSs obtained from different EDF combinations without additional modification of the density-dependent three-body parameters.
    }
    \label{fig:eos_branch_three_panels}
\end{figure}

The three cases are illustrated in Fig.~\ref{fig:eos_branch_three_panels}.
In curve (i), the calculated EOS branch is monotonic in the $P$-$\varepsilon$ plane.
In curve (ii), the branch develops a single extremum and then enters a negative-slope segment.
In curve (iii), the branch develops two extrema, so that the negative-slope segment is bounded by two monotonic branches.
We treat these three cases separately as follows.

(i) A monotonic EOS branch has no extremum,
\begin{align}
N_{\rm ext}=0 .
\end{align}
This case corresponds to Fig.~\ref{fig:eos_branch_three_panels}(i).
Here $P$ increases monotonically with $\varepsilon$, and the EOS can be used directly in the TOV integration.
The equilibrium sequence then provides an unambiguous energy density for each pressure encountered during the integration.

(ii) A one-extremum EOS branch has
\begin{align}
N_{\rm ext}=1 .
\end{align}
This case corresponds to Fig.~\ref{fig:eos_branch_three_panels}(ii).
At the extremum, $dP/d\varepsilon=0$, and the continuation beyond it contains a segment with
\begin{align}
\frac{\dd P}{\dd \varepsilon}
<
0 .
\label{eq:one_extremum_negative_slope}
\end{align}
For a cold barotropic EOS, $\dd P/\dd \varepsilon$ corresponds to the squared sound speed in units $c=1$; therefore, a negative value indicates mechanical instability.
In the present treatment, such an EOS is used only up to the monotonic branch before the first extremum.
Configurations whose central pressure would require the continuation along the mechanically unstable segment are not included in the TOV sequence.
This case is treated as a branch-limited EOS, but is not identified by itself as a first-order coexistence region, because it does not provide two stable endpoints to be connected by a phase-construction procedure.

(iii) A two-extrema EOS branch has
\begin{align}
N_{\rm ext}=2 .
\end{align}
This case corresponds to Fig.~\ref{fig:eos_branch_three_panels}(iii).
In this case, the calculated $P(\varepsilon)$ sequence contains an intermediate negative-slope segment,
\begin{align}
\frac{\dd P}{\dd \varepsilon}
<
0 ,
\label{eq:two_extrema_negative_slope}
\end{align}
bounded by two monotonic branches.
Such a structure is treated as a Maxwell-construction candidate, because it may indicate the appearance of a first-order coexistence region.
However, the existence of two extrema alone is not taken as a complete phase-construction prescription.
Whether the intermediate segment should be replaced by a Maxwell construction is determined by the coexistence conditions discussed in the next subsection.

This classification is defined only by the monotonicity of the calculated $P(\varepsilon)$ sequence and by the number of extrema.
The one-extremum case is treated as a termination of the directly usable EOS branch, whereas the two-extrema case is treated as a candidate first-order coexistence region.
The classification is a practical treatment of the EOS branches used for neutron-star structure calculations, not an additional physical assumption about the underlying interaction.

In the parameter-scan results presented in Secs.~\ref{sec:lln_scan} and \ref{sec:lnn_scan}, we further distinguish two extremum counts.
The quantity $N_{\rm ext}^{\rm full}$ denotes the number of extrema in the full calculated EOS sequence.
The quantity $N_{\rm ext}^{\rm NS}$ counts only extrema encountered within the pressure range relevant to the neutron-star sequence up to the configuration used to determine the maximum mass.
This distinction is useful because extrema may appear only in the ultra-high-density part of the calculated EOS, beyond the central pressure reached by the maximum-mass configuration.
The Maxwell construction used for the two-extrema case is described in the next subsection.

\subsection{Maxwell construction for nonmonotonic EOSs}
\label{sec:maxwell_construction}

For a monotonic EOS branch, the beta-equilibrated sequence directly provides a single-valued relation $P(\varepsilon)$ for the TOV calculation. 
When the calculated sequence contains a negative-slope segment, however, the direct branch cannot be used as a mechanically admissible EOS. 
For a cold barotropic EOS, $\dd P/\dd \varepsilon$ gives the squared sound speed in units $c=1$, and a segment with $dP/d\varepsilon<0$ corresponds to negative compressibility. 
In the two-extrema case discussed in Sec.~\ref{sec:eos_branches}, this negative-slope segment is bounded by two monotonic branches. 
We treat such a case as a Maxwell-construction candidate.

The Maxwell construction used here is based on the equality of pressure and baryon chemical potential between the low-density and high-density phases. 
We first show why the baryon chemical potential along the charge-neutral beta-equilibrated branch is equal to the neutron chemical potential. 
At zero temperature, the differential of the energy density is
\begin{equation}
    \dd \varepsilon
    =
    \mu_n \dd n_n
    +\mu_p \dd n_p
    +\mu_\Lambda \dd n_\Lambda
    +\mu_e \dd n_e
    +\mu_\mu \dd n_\mu .
    \label{eq:de_full}
\end{equation}
The beta-equilibrium conditions are
\begin{equation}
    \mu_n=\mu_p+\mu_e,\qquad
    \mu_e=\mu_\mu,\qquad
    \mu_n=\mu_\Lambda .
    \label{eq:beta_equilibrium_maxwell}
\end{equation}
Using these relations, Eq.~\eqref{eq:de_full} becomes
\begin{align}
    \dd \varepsilon
    &=
    \mu_n \dd n_n
    +(\mu_n-\mu_e)\dd n_p
    +\mu_n \dd n_\Lambda
    +\mu_e \dd n_e
    +\mu_e \dd n_\mu
    \nonumber\\
    &=
    \mu_n(\dd n_n+\dd n_p+\dd n_\Lambda)
    +\mu_e(\dd n_e+\dd n_\mu-\dd n_p).
    \label{eq:de_rearranged}
\end{align}
The total baryon density and local charge neutrality are
\begin{equation}
    n_B=n_n+n_p+n_\Lambda,\qquad
    n_p=n_e+n_\mu .
    \label{eq:baryon_charge_constraints_maxwell}
\end{equation}
Therefore,
\begin{equation}
    \dd n_B=\dd n_n+\dd n_p+\dd n_\Lambda,\qquad
    \dd n_p=\dd n_e+\dd n_\mu .
    \label{eq:dbaryon_dcharge_constraints}
\end{equation}
Substituting Eq.~\eqref{eq:dbaryon_dcharge_constraints} into Eq.~\eqref{eq:de_rearranged}, we obtain
\begin{equation}
    \dd\varepsilon=\mu_n \dd n_B .
    \label{eq:de_muB}
\end{equation}
Thus, along the locally charge-neutral beta-equilibrated branch, the baryon chemical potential is
\begin{equation}
    \mu_B
    \equiv
    \frac{\dd\varepsilon}{\dd n_B}
    =
    \mu_n .
    \label{eq:maxwell_muB}
\end{equation}

The same result is consistent with the thermodynamic expression for the pressure. 
Using
\begin{equation}
    P=\sum_{i=n,p,\Lambda,e,\mu} n_i\mu_i-\varepsilon ,
    \label{eq:pressure_maxwell}
\end{equation}
and the beta-equilibrium relations in Eq.~\eqref{eq:beta_equilibrium_maxwell}, one obtains
\begin{align}
    \sum_i n_i\mu_i
    &=
    \mu_n(n_n+n_p+n_\Lambda)
    +\mu_e(n_e+n_\mu-n_p)
    \nonumber\\
    &=
    \mu_n n_B ,
    \label{eq:gibbs_per_baryon_reduction}
\end{align}
where local charge neutrality has been used in the last step. 
Therefore,
\begin{equation}
    P=n_B\mu_n-\varepsilon,
    \qquad
    \mu_B=\mu_n=\frac{\varepsilon+P}{n_B}.
    \label{eq:muB_pressure_relation}
\end{equation}

For a Maxwell construction, we search for two densities $n_B^L$ and $n_B^H$ on the low-density and high-density monotonic branches such that
\begin{equation}
    P(n_B^L)=P(n_B^H)\equiv P_M ,
    \label{eq:maxwell_pressure}
\end{equation}
and
\begin{equation}
    \mu_B(n_B^L)=\mu_B(n_B^H).
    \label{eq:maxwell_muB_equal}
\end{equation}
Using Eq.~\eqref{eq:maxwell_muB}, the second condition is written in the present calculation as
\begin{equation}
    \mu_n(n_B^L)=\mu_n(n_B^H).
    \label{eq:maxwell_mun_equal}
\end{equation}
Equations~\eqref{eq:maxwell_pressure} and \eqref{eq:maxwell_mun_equal} are the coexistence conditions used to replace the nonmonotonic part of the EOS.

Once the two endpoints are found, the EOS between them is replaced by a constant-pressure coexistence segment,
\begin{equation}
    P(\varepsilon)=P_M,
    \qquad
    \varepsilon_L \leq \varepsilon \leq \varepsilon_H ,
    \label{eq:maxwell_plateau}
\end{equation}
where
\begin{equation}
    \varepsilon_L=\varepsilon(n_B^L),\qquad
    \varepsilon_H=\varepsilon(n_B^H).
    \label{eq:maxwell_energy_endpoints}
\end{equation}
Equivalently, the energy density jumps from $\varepsilon_L$ to $\varepsilon_H$ at the same pressure $P_M$. 
This construction removes the mechanically unstable negative-slope segment from the barotropic EOS used in the TOV calculation.

The present Maxwell construction should be distinguished from a Gibbs construction \cite{glendenning1992firstorder}.
In the Maxwell treatment adopted here, local charge neutrality is imposed separately on each branch, and the coexistence condition is formulated in terms of a single baryon chemical potential $\mu_B$. 
A Gibbs construction would instead allow the two phases to carry different local charge densities while enforcing global charge neutrality, requiring the simultaneous equality of pressure, baryon chemical potential, and charge chemical potential. 
Such an extension may give a smoother mixed-phase EOS, but it is beyond the scope of the present deterministic analysis.

In the results below, a two-extrema EOS is therefore treated as a Maxwell-construction candidate. 
A one-extremum EOS is not subjected to the Maxwell construction, because it does not provide two monotonic branches that can serve as the low-density and high-density endpoints of a coexistence construction. 
Instead, it is treated as a branch-limited EOS and is used only up to the first extremum.

\section{Deterministic effects of hyperonic density-dependent three-body terms on neutron-star EOSs and structure}
\label{sec:deterministic_effect}
In this section, we present deterministic results for neutron-star EOSs and mass--radius relations obtained by varying the hyperonic density-dependent three-body terms.
Here, the $\Lambda\Lambda N$ and $\Lambda NN$ density-dependent terms denote the density-dependent contributions introduced in the $\Lambda\Lambda$ and $N\Lambda$ EDF sectors, respectively.
The two terms are varied separately in order to isolate their effects on the EOS, the mass--radius relation, and the maximum mass.
Bayesian and machine-learning analyses of the interaction parameters are presented separately in Sec.~\ref{sec:bayesian_ml}.
\subsection{Computational details}
\label{sec:computational_details}

For each point in the $(\gamma,C_3)$ or $(\beta,A_3)$ parameter plane, we construct a beta-equilibrated EOS and solve the corresponding TOV sequence. 
The density-dependent parameters are updated at the EDF level, while the remaining nucleonic and hyperonic interaction parameters are kept fixed to the reference set specified for each scan. 
Thus, one parameter point corresponds to one tabulated EOS,
\begin{equation}
    n_B \mapsto [\varepsilon(n_B),P(n_B)],
\end{equation}
and to one mass--radius sequence.

The EOS is constructed by solving the equilibrium conditions at fixed total baryon density $n_B$. 
When all particle species are present, the unknown density vector is
\begin{equation}
    \boldsymbol{x}
    =
    (n_n,n_p,n_\Lambda,n_e,n_\mu),
\end{equation}
and the nonlinear equations are
\begin{align}
    F_1 &= n_n+n_p+n_\Lambda-n_B =0, \\
    F_2 &= n_p-n_e-n_\mu =0, \\
    F_3 &= \mu_n-\mu_p-\mu_e =0, \\
    F_4 &= \mu_e-\mu_\mu =0, \\
    F_5 &= \mu_n-\mu_\Lambda =0 .
\end{align}
These equations express baryon-number conservation, charge neutrality, and beta equilibrium. 
They are solved by Newton iteration at each prescribed value of $n_B$. 
The solution at the preceding density point is used as the initial guess for the next point, so that the EOS is followed as a continuous sequence in $n_B$. 
If a particle species is absent, its density is fixed to zero and the corresponding equality condition is replaced by the appropriate onset condition. 
In particular, below the $\Lambda$-onset density we impose $n_\Lambda=0$ and monitor the threshold condition discussed in Sec.~\ref{sec:onset_condition}.

After the equilibrium composition is obtained, the total energy density and pressure are evaluated from the EDF and the thermodynamic relation in Sec.~\ref{sec:skyrme_edf_hyperonic_ns}. 
This gives the tabulated EOS sequence $n_B\mapsto[\varepsilon(n_B),P(n_B)]$. 
The same sequence is inspected for extrema in the $P$-$\varepsilon$ plane. 
The number of extrema in the full calculated sequence defines $N_{\rm ext}^{\rm full}$. 
For the TOV calculation, only the mechanically admissible part of the EOS is used. 
If the calculated branch reaches an extremum and no larger pressure can be referenced along the admissible branch, the mass--radius sequence is terminated at that point. 
Such a case is classified as a branch-limited EOS, as described in Sec.~\ref{sec:eos_branches}.

The mass--radius relation is obtained by solving the TOV equations as an initial-value problem for a sequence of central pressures $P_c$. 
In the deterministic scans, the central pressure $P_c$ is sampled over the admissible pressure range of each EOS. 
The scan is not continued beyond the pressure range for which the EOS provides a mechanically admissible input to the TOV equations. 
For each $P_c$, the tabulated EOS is linearly interpolated to obtain the corresponding energy density $\varepsilon(P_c)$. 
During the TOV integration, the same linear interpolation is used to evaluate $\varepsilon(P)$ at the local pressure. 
The TOV equations are integrated outward from the stellar center using the fourth-order Runge--Kutta method until the pressure reaches zero. 
The radius $R$ is defined by $P(R)=0$, and the gravitational mass is $M=m(R)$. 
Thus, each value of $P_c$ gives one point $(R,M)$ in the mass--radius plane.

The maximum mass $M_{\max}$ is extracted from the resulting TOV sequence. 
For each EOS, the scan in $P_c$ is stopped when the causal boundary $dP/d\varepsilon=1$ is reached, when the next trial central pressure gives a smaller gravitational mass, or when the required pressure exceeds the admissible pressure range of the EOS table. 
The last criterion is relevant for branch-limited EOSs, where the usable EOS ends at the first extremum.

For monotonic EOSs, this procedure gives the ordinary maximum mass along the directly constructed EOS branch. 
For one-extremum EOSs, the TOV sequence is constructed only up to the largest admissible pressure before the mechanically unstable continuation. 
For two-extrema EOSs, the branch is identified as a Maxwell-construction candidate. 
In the deterministic maps, we do not construct a Maxwell-corrected $M_{\max}$ over the full parameter plane; instead, $M_{\max}$, $N_{\rm ext}^{\rm NS}$, and $N_{\rm ext}^{\rm full}$ are used together to diagnose the mass response and the admissibility of the EOS branch.

\subsection{Impact of the \texorpdfstring{$\Lambda\Lambda N$}{LambdaLambdaN} density-dependent term on neutron-star EOSs and structure}
\label{sec:lln_scan}

We first examine the density-dependent $\Lambda\Lambda N$ contribution in the $\Lambda\Lambda$ sector. In the EDF convention adopted in this work, this contribution is written as
\begin{equation}
    \varepsilon_{\Lambda\Lambda N}
    =
    C_3 n_\Lambda^2 n_N^\gamma ,
    \label{eq:lln_edf}
\end{equation}
where $n_N=n_n+n_p$. The coefficient $C_3$ is the EDF-level strength of the density-dependent $\Lambda\Lambda N$ term, and $\gamma$ controls its density dependence. In this subsection, the $\Lambda NN$ density-dependent term is kept fixed, and the two $\Lambda\Lambda N$ parameters $(\gamma,C_3)$ are varied in order to isolate the response of the EOS and neutron-star structure to the $\Lambda\Lambda N$ term.

\begin{figure}[t]
    \centering
    \includegraphics[width=\columnwidth]{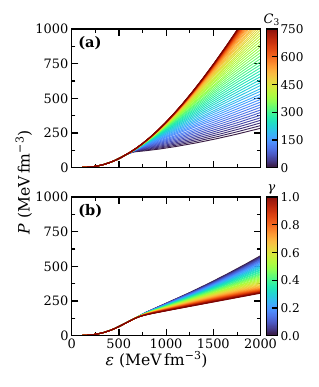}
    \caption{
    Representative EOS bands showing the response to the $\Lambda\Lambda N$ density-dependent term.
    Panel (a) shows the EOSs obtained by varying $C_3$ at fixed $\gamma=1/3$, while panel (b) shows the EOSs obtained by varying $\gamma$ at fixed $C_3=125$.
    The curves remain close to each other before a finite $\Lambda$ fraction is generated, whereas they separate in the post-onset region.
    Increasing $C_3$ enhances the repulsive finite-$\Lambda$ contribution and raises the post-onset pressure.
    At fixed $C_3$, increasing $\gamma$ tends to reduce this pressure enhancement in the neutron-star-relevant density range.
    }
    \label{fig:lln_eos_band}
\end{figure}

Before discussing the two-dimensional parameter maps, it is useful to inspect the EOS response itself. Figure~\ref{fig:lln_eos_band} shows representative EOS bands obtained by varying one $\Lambda\Lambda N$ parameter at a time. In panel~(a), $C_3$ is varied at fixed $\gamma=1/3$. In panel~(b), $\gamma$ is varied at fixed $C_3=125$. These bands are not intended to define an additional parameter constraint; rather, they illustrate how the $\Lambda\Lambda N$ term changes the pressure--energy-density relation before the corresponding TOV calculation is performed.

The behavior shown in Fig.~\ref{fig:lln_eos_band} follows directly from the structure of Eq.~\eqref{eq:lln_edf}. Since the $\Lambda\Lambda N$ term is proportional to $n_\Lambda^2$, its effect is absent before a finite $\Lambda$ fraction is produced. Once $\Lambda$ hyperons appear, however, the term contributes to the chemical potentials and pressure. Increasing $C_3$ therefore increases the finite-$\Lambda$ repulsion and generally stiffens the post-onset hyperonic EOS relative to the baseline hyperonic case. By contrast, the density exponent $\gamma$ controls how this repulsive contribution is distributed over density and composition. In the present scan, larger $\gamma$ tends to weaken the pressure enhancement in the neutron-star-relevant density range. This trend becomes less visible when $C_3$ is small, because the overall $\Lambda\Lambda N$ contribution itself is then small.

The same structure also explains why the $\Lambda\Lambda N$ term is onset-inactive in the specific sense used in this work. As shown in Sec.~\ref{sec:onset_condition}, the term is proportional to $n_\Lambda^2$, and therefore its first derivative with respect to $n_\Lambda$ vanishes at $n_\Lambda=0$. For a fixed $N\Lambda$ sector, varying $C_3$ or $\gamma$ does not directly shift the local continuous $\Lambda$-onset condition. This statement should be understood only as a threshold statement. After the onset, the same term modifies the finite-$\Lambda$ EOS and can change the pressure response, the maximum mass, and the stability range of the stellar branch.

\begin{figure*}[t]
    \centering
    \includegraphics[width=\textwidth]{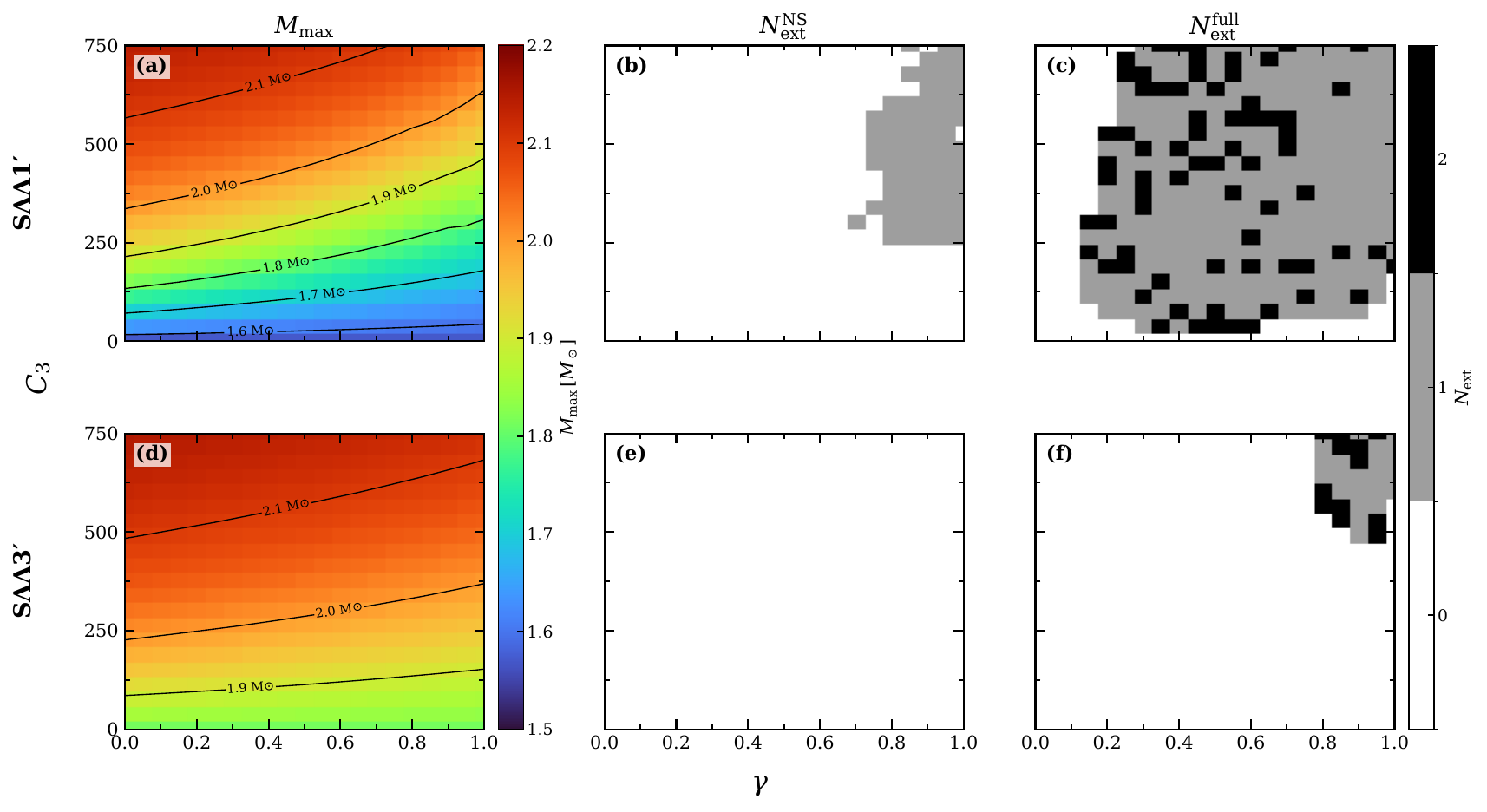}
    \caption{
Maps of the $\Lambda\Lambda N$ density-dependent scan in the $(\gamma,C_3)$ plane.
The NN and $N\Lambda$ sectors are fixed to BSk24 and SLL4$^\prime$, respectively, and only the $\Lambda\Lambda$ reference interaction is changed between the two rows.
The density-dependent $\Lambda\Lambda N$ term is absent in the fitted baseline, corresponding to $(\gamma,C_3)=(-,0)$; here $\gamma$ is irrelevant when $C_3=0$.
The coefficient $C_3$ is quoted in units of ${\rm MeV\,fm}^{3\gamma+3}$.
Panels (a)--(c) correspond to the S$\Lambda\Lambda$1$^\prime$ reference interaction, and panels (d)--(f) correspond to the S$\Lambda\Lambda$3$^\prime$ reference interaction.
Panels (a) and (d) show the maximum gravitational mass $M_{\max}$, panels (b) and (e) show the number of extrema in the neutron-star-relevant EOS branch, $N_{\rm ext}^{\rm NS}$, and panels (c) and (f) show the number of extrema in the full calculated EOS, $N_{\rm ext}^{\rm full}$.
Black contours in the $M_{\max}$ panels denote constant values of $M_{\max}$ in units of $M_\odot$.
White contours denote lower-mass contours at and below $1.5M_\odot$.
    }
    \label{fig:lln_mmax_extremum_count}
\end{figure*}

The full two-dimensional scan is summarized in Fig.~\ref{fig:lln_mmax_extremum_count}. Panels (a)--(c) show the $S\Lambda\Lambda1'$ reference interaction, and panels (d)--(f) show the $S\Lambda\Lambda3'$ reference interaction. The left column, panels (a) and (d), shows the maximum gravitational mass $M_{\max}$; the middle column, panels (b) and (e), shows $N_{\rm ext}^{\rm NS}$; and the right column, panels (c) and (f), shows $N_{\rm ext}^{\rm full}$. The mass maps should be read together with the extremum-count maps, because the same value of $M_{\max}$ can have different physical interpretations depending on whether the EOS branch remains mechanically admissible up to the central pressure relevant for the maximum-mass configuration.

Here $N_{\rm ext}^{\rm NS}$ and $N_{\rm ext}^{\rm full}$ diagnose different density ranges of the same calculated EOS. The quantity $N_{\rm ext}^{\rm NS}$ counts extrema encountered by the neutron-star branch up to the configuration used to determine $M_{\max}$. Thus, $N_{\rm ext}^{\rm NS}=0$ means that the stellar branch remains mechanically admissible, whereas $N_{\rm ext}^{\rm NS}=1$ means that the branch reaches an extremum and its continuation contains a negative-slope segment with $dP/d\varepsilon<0$. As discussed in Sec.~\ref{sec:eos_branches}, this continuation is mechanically unstable and is not used in the TOV integration.

By contrast, $N_{\rm ext}^{\rm full}$ counts extrema in the full calculated EOS sequence, including the high-density continuation beyond the pressure range sampled by the maximum-mass star. Therefore, $N_{\rm ext}^{\rm full}>0$ with $N_{\rm ext}^{\rm NS}=0$ indicates nonmonotonic behavior only in the ultra-high-density continuation, not an instability of the neutron-star branch used in the present TOV calculation.

This distinction is important for interpreting the $\Lambda\Lambda N$ scan. In the $S\Lambda\Lambda1'$ case, the full-EOS extremum map in Fig.~\ref{fig:lln_mmax_extremum_count}(c) shows that extrema appear over a broad region. However, the corresponding neutron-star-relevant extremum map in Fig.~\ref{fig:lln_mmax_extremum_count}(b) shows that many of these extrema are not encountered by the neutron-star branch before the maximum mass is determined. These regions should therefore not be interpreted as instabilities of the neutron-star configurations that determine $M_{\max}$.

A smaller part of the $S\Lambda\Lambda1'$ parameter plane has $N_{\rm ext}^{\rm NS}=1$, as shown in Fig.~\ref{fig:lln_mmax_extremum_count}(b). In this region, the neutron-star-relevant branch itself reaches an extremum. Consequently, the continuation of the EOS would enter a mechanically unstable segment, and the stellar sequence is limited by the first extremum. This branch limitation affects the appearance of the mass contours in Fig.~\ref{fig:lln_mmax_extremum_count}(a). In particular, the otherwise smooth variation of $M_{\max}$ can become less smooth or piecewise in the region where the stellar branch terminates.

The $S\Lambda\Lambda3'$ case shows a more stable behavior in the neutron-star-relevant region. As shown in Fig.~\ref{fig:lln_mmax_extremum_count}(e), the $N_{\rm ext}^{\rm NS}$ map is almost entirely zero, indicating that the EOS branch used to construct the neutron-star sequence remains mechanically admissible throughout nearly the whole scan. Although a small number of extrema may still appear in the full-EOS extremum map, Fig.~\ref{fig:lln_mmax_extremum_count}(f), they occur only in the high-density continuation of the EOS and do not affect the maximum-mass stellar branch. Thus, for $S\Lambda\Lambda3'$, the $M_{\max}$ map in Fig.~\ref{fig:lln_mmax_extremum_count}(d) can be interpreted more directly as the response of a mechanically admissible neutron-star branch.

Another feature of the $\Lambda\Lambda N$ scan is that the regions with $N_{\rm ext}^{\rm full}=2$, visible in the full-EOS extremum maps in Figs.~\ref{fig:lln_mmax_extremum_count}(c) and \ref{fig:lln_mmax_extremum_count}(f), do not form a simple connected domain. In the present results, two-extrema structures in the full EOS appear in a scattered manner, especially compared with the more organized branch structure found in the $\Lambda NN$ scan discussed below. This suggests that, for the $\Lambda\Lambda N$ term, the second extremum in the ultra-high-density continuation is not a robust feature controlled by a single simple boundary in the $(\gamma,C_3)$ plane.

We now turn to the maximum-mass panels, Figs.~\ref{fig:lln_mmax_extremum_count}(a) and \ref{fig:lln_mmax_extremum_count}(d). The dominant trend is along the $C_3$ direction. For both $S\Lambda\Lambda1'$ and $S\Lambda\Lambda3'$, increasing $C_3$ raises $M_{\max}$, reflecting the enhanced repulsive finite-$\Lambda$ contribution of the $\Lambda\Lambda N$ term. This behavior is consistent with the EOS response shown in Fig.~\ref{fig:lln_eos_band}, where increasing $C_3$ increases the post-onset pressure relative to the baseline curve.

The role of $\gamma$ is seen in the slope of the $M_{\max}$ contours. At fixed $C_3$, increasing $\gamma$ tends to reduce the mass recovery in the neutron-star-relevant density range. Equivalently, a larger $C_3$ is required to maintain the same value of $M_{\max}$ at larger $\gamma$. This trend follows the EOS-band behavior at fixed $C_3$, where increasing $\gamma$ weakens the pressure enhancement produced by the $\Lambda\Lambda N$ term. Therefore, $\gamma$ should be interpreted as controlling the density dependence and effectiveness of the finite-$\Lambda$ repulsion, rather than as an independent source of mass increase.

The $2M_\odot$ contour in the maximum-mass panels, Figs.~\ref{fig:lln_mmax_extremum_count}(a) and \ref{fig:lln_mmax_extremum_count}(d), is used here as a visual benchmark for mass recovery, motivated by the two-solar-mass constraint associated with the hyperon puzzle. For $S\Lambda\Lambda1'$, a relatively large $C_3$ is required to recover $M_{\max}$ to the $2M_\odot$ level. For $S\Lambda\Lambda3'$, the same contour appears at lower $C_3$, indicating that the baseline $\Lambda\Lambda$ reference interaction already gives a larger maximum mass before the additional $\Lambda\Lambda N$ repulsion is varied.

The mass contours in Figs.~\ref{fig:lln_mmax_extremum_count}(a) and \ref{fig:lln_mmax_extremum_count}(d) must also be read together with the corresponding $N_{\rm ext}^{\rm NS}$ maps in Figs.~\ref{fig:lln_mmax_extremum_count}(b) and \ref{fig:lln_mmax_extremum_count}(e). In the $S\Lambda\Lambda1'$ case, part of the high-$\gamma$, high-$C_3$ region has $N_{\rm ext}^{\rm NS}=1$, as shown in Fig.~\ref{fig:lln_mmax_extremum_count}(b). There, the neutron-star-relevant EOS branch reaches an extremum, and the maximum mass is influenced by the termination of the mechanically admissible branch. Consequently, the less smooth or more piecewise appearance of the contours in this region should not be interpreted as a purely smooth stiffening effect.

The present $\Lambda\Lambda N$ scan therefore leads to three conclusions. First, the $\Lambda\Lambda N$ density-dependent term is onset-inactive at the level of the local continuous threshold condition, but it can still have a sizable effect on the finite-$\Lambda$ EOS after the onset. Second, the distinction between $N_{\rm ext}^{\rm NS}$ and $N_{\rm ext}^{\rm full}$ is essential: an extremum in the full high-density continuation does not necessarily imply an instability of the neutron-star branch. Third, the recovery of $M_{\max}$ should be discussed through the $M_{\max}$ contour pattern together with $N_{\rm ext}^{\rm NS}$, rather than through the largest mass value obtained at the boundary of the scan. The comparison with the $\Lambda NN$ scan below will clarify how this finite-$\Lambda$ response differs from the onset-active response in the $N\Lambda$ sector.

\subsection{Impact of the \texorpdfstring{$\Lambda NN$}{LambdaNN} density-dependent term on neutron-star EOSs and structure}
\label{sec:lnn_scan}

We next examine the density-dependent $\Lambda NN$ contribution in the $N\Lambda$ sector. In the EDF convention adopted in this work, this contribution is written as
\begin{equation}
    \varepsilon_{\Lambda NN}
    =
    A_3 n_\Lambda n_N^{1+\beta},
    \label{eq:lnn_edf}
\end{equation}
where $n_N=n_n+n_p$. The coefficient $A_3$ is the EDF-level strength of the density-dependent $\Lambda NN$ term, and $\beta$ controls its density dependence. In this subsection, the $\Lambda\Lambda N$ density-dependent term is kept fixed, and the two $\Lambda NN$ parameters $(\beta,A_3)$ are varied in order to isolate the response of the EOS and neutron-star structure to the $\Lambda NN$ term.

\begin{figure}[t]
    \centering
    \includegraphics[width=\columnwidth]{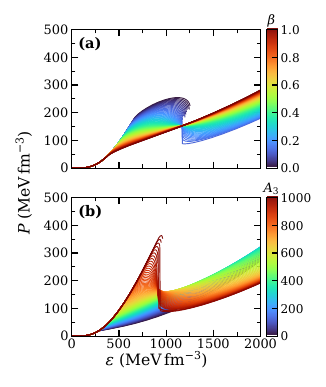}
    \caption{
    Representative EOS bands showing the response to the $\Lambda NN$ density-dependent term.
    Panel (a) shows the EOSs obtained by varying $\beta$ at fixed $A_3=1000$, while panel (b) shows the EOSs obtained by varying $A_3$ at fixed $\beta=1/3$.
    Since the $\Lambda NN$ term is linear in $n_\Lambda$, it can modify both the $\Lambda$-onset condition and the post-onset pressure response.
    The apparent concentration of the curves in panel (a) should not be interpreted as an exact fixed point of the EOS; it reflects a partial compensation between the explicit density dependence and the equilibrium composition response in this density range.
    }
    \label{fig:lnn_eos_band}
\end{figure}

Before discussing the two-dimensional parameter maps, it is useful to inspect the EOS response itself. Figure~\ref{fig:lnn_eos_band} shows representative EOS bands obtained by varying one $\Lambda NN$ parameter at a time. In panel~(a), $\beta$ is varied at fixed $A_3=1000$. In panel~(b), $A_3$ is varied at fixed $\beta=1/3$. Unlike the $\Lambda\Lambda N$ case, the $\Lambda NN$ term can modify the $\Lambda$-onset condition itself. Therefore, the separation, curvature change, and possible nonmonotonic behavior of the EOS curves reflect both the shift of the $\Lambda$-onset point and the modification of the post-onset pressure response.

The qualitative difference from the $\Lambda\Lambda N$ scan follows directly from the structure of Eq.~\eqref{eq:lnn_edf}. As shown in Sec.~\ref{sec:onset_condition}, the $\Lambda NN$ term is linear in $n_\Lambda$, and its first derivative with respect to $n_\Lambda$ remains finite at $n_\Lambda=0$. Varying $A_3$ or $\beta$ can therefore directly change the local continuous $\Lambda$-onset condition. This onset-active character is essential for interpreting the results below: the $\Lambda NN$ scan modifies not only the finite-$\Lambda$ EOS after the onset, but also the density at which the hyperonic branch begins.

\begin{figure*}[t]
    \centering
    \includegraphics[width=\textwidth]{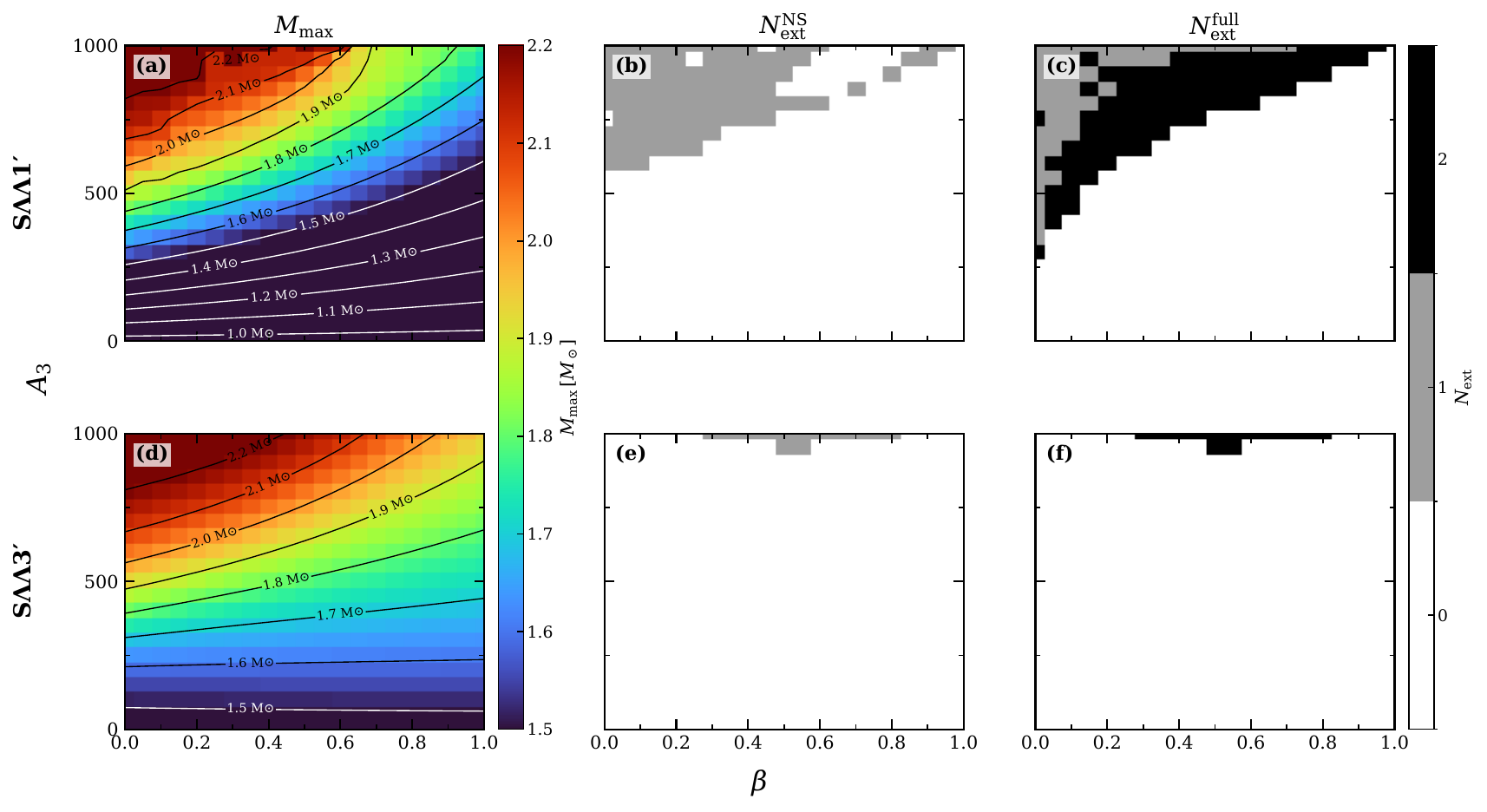}
    \caption{
Maps of the $\Lambda NN$ density-dependent scan in the $(\beta,A_3)$ plane.
The NN and $N\Lambda$ sectors are fixed to BSk24 and SLL4$^\prime$, respectively, and only the $\Lambda\Lambda$ reference interaction is changed between the two rows.
The fitted baseline value of the density-dependent $\Lambda NN$ term is $(\beta,A_3)=(1,705)$.
The coefficient $A_3$ is quoted in units of ${\rm MeV\,fm}^{3\beta+3}$.
Panels (a)--(c) correspond to the S$\Lambda\Lambda$1$^\prime$ reference interaction, and panels (d)--(f) correspond to the S$\Lambda\Lambda$3$^\prime$ reference interaction.
Panels (a) and (d) show the maximum gravitational mass $M_{\max}$, panels (b) and (e) show the number of extrema in the neutron-star-relevant EOS branch, $N_{\rm ext}^{\rm NS}$, and panels (c) and (f) show the number of extrema in the full calculated EOS, $N_{\rm ext}^{\rm full}$.
Black contours in the $M_{\max}$ panels denote constant values of $M_{\max}$ in units of $M_\odot$.
White contours denote lower-mass contours at and below $1.5M_\odot$.
    }
    \label{fig:lnn_mmax_extremum_count}
\end{figure*}

The full two-dimensional scan is summarized in Fig.~\ref{fig:lnn_mmax_extremum_count}. Panels (a)--(c) show the $S\Lambda\Lambda1'$ reference interaction, and panels (d)--(f) show the $S\Lambda\Lambda3'$ reference interaction. The left column, panels (a) and (d), shows the maximum gravitational mass $M_{\max}$; the middle column, panels (b) and (e), shows $N_{\rm ext}^{\rm NS}$; and the right column, panels (c) and (f), shows $N_{\rm ext}^{\rm full}$. We use the same interpretation of $N_{\rm ext}^{\rm NS}$ and $N_{\rm ext}^{\rm full}$ as in Sec.~\ref{sec:lln_scan}. In the present $\Lambda NN$ scan, however, the extremum structure is more directly tied to the onset-active modification of the hyperonic branch.

The extremum-count maps show a clear contrast with the $\Lambda\Lambda N$ scan. In the $S\Lambda\Lambda1'$ case, the nonmonotonic region forms an organized domain in the $(\beta,A_3)$ plane, as seen in Figs.~\ref{fig:lnn_mmax_extremum_count}(b) and \ref{fig:lnn_mmax_extremum_count}(c), rather than appearing as scattered isolated points. The two-extrema region in Fig.~\ref{fig:lnn_mmax_extremum_count}(c) is interpreted as a Maxwell-construction candidate, because the negative-slope segment is bounded by two monotonic branches. By contrast, the one-extremum branch-limited region in Fig.~\ref{fig:lnn_mmax_extremum_count}(b) does not reconnect to a mechanically admissible high-density branch within the calculated density range. We do not attempt to rank the degree of instability within these nonmonotonic branches in the present work; instead, we use the count $N_{\rm ext}^{\rm NS}$ as an operational classification of whether the branch remains usable for the TOV sequence up to $M_{\max}$.

The $S\Lambda\Lambda3'$ case is more stable, but the same mechanism is still visible in the extremum-count maps, Figs.~\ref{fig:lnn_mmax_extremum_count}(e) and \ref{fig:lnn_mmax_extremum_count}(f). The region with $N_{\rm ext}^{\rm NS}=1$, shown in Fig.~\ref{fig:lnn_mmax_extremum_count}(e), is much smaller than in the $S\Lambda\Lambda1'$ case, and most of the parameter plane remains mechanically admissible for the neutron-star branch. Nevertheless, nonmonotonic behavior is not completely absent, as indicated by the full-EOS extremum map in Fig.~\ref{fig:lnn_mmax_extremum_count}(f). This shows that the branch structure is controlled not only by the $\Lambda NN$ term itself, but also by the underlying $\Lambda\Lambda$ reference interaction, which determines the finite-$\Lambda$ response after the onset.

The two-extrema regions in the full-EOS extremum maps, Figs.~\ref{fig:lnn_mmax_extremum_count}(c) and \ref{fig:lnn_mmax_extremum_count}(f), are treated as Maxwell-construction candidates. We do not construct a Maxwell-corrected maximum-mass map over the full $(\beta,A_3)$ plane in this work. Instead, the $M_{\max}$ maps in Figs.~\ref{fig:lnn_mmax_extremum_count}(a) and \ref{fig:lnn_mmax_extremum_count}(d) are used to diagnose how the directly constructed EOS sequence or its branch-limited segment responds across the parameter plane. In the Maxwell-candidate region, the plotted value of $M_{\max}$ should therefore not be interpreted as the final maximum mass after a complete phase-construction treatment.

\begin{figure}[t]
    \centering
    \includegraphics[width=1.09\columnwidth]{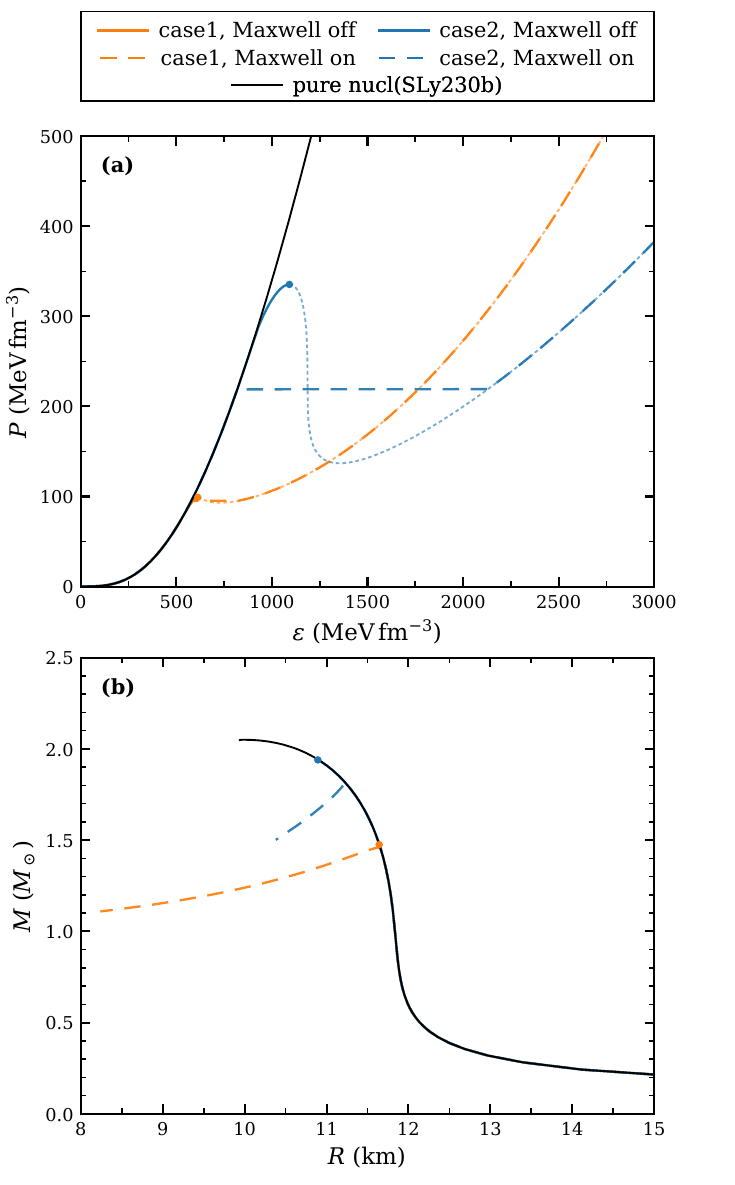}\hspace{-5mm}\vspace{-5mm}
    \caption{
    Representative Maxwell construction and the corresponding mass--radius response in the $\Lambda NN$ scan.
    Panel (a) shows the EOS in the $P$--$\varepsilon$ plane, and panel (b) shows the corresponding mass--radius sequences.
    Solid curves denote the EOS segments used in the stellar-structure calculation, whereas dotted curves indicate mechanically unstable parts of the original nonmonotonic EOS before the Maxwell replacement.
    The Maxwell construction replaces these segments by constant-pressure coexistence segments.
    Markers indicate the connection points between the admissible branch and the nonmonotonic or coexistence segment.
    Case 1 illustrates a less extreme Maxwell-corrected sequence, where the mass still increases slightly after the transition.
    Case 2, obtained for the SLy230b+SLL4$^{\prime}$+S$\Lambda\Lambda1$ interaction with $(A_3,\beta)=(800,0.25)$, bends downward almost immediately after the onset or coexistence point, so that $M_{\max}$ is effectively onset-controlled.
    }
    \label{fig:lnn_maxwell_mr}
\end{figure}

To illustrate the qualitative consequence of the Maxwell construction, we select two representative parameter points from the two-extrema region. 
Figure~\ref{fig:lnn_maxwell_mr} compares the EOSs before and after the Maxwell construction and shows the corresponding mass--radius sequences. 
In the EOS panel, the solid curves denote the EOS segments used in the stellar-structure calculation, while the dotted segments indicate the mechanically unstable parts of the original nonmonotonic EOS before the Maxwell replacement. 
The Maxwell-constructed EOS replaces these dotted segments by constant-pressure coexistence segments. 
The markers indicate the connection points between the admissible EOS branch and the nonmonotonic or coexistence segment; the same markers are shown on the corresponding mass--radius curves in panel~(b). 
The two examples demonstrate that Maxwell-corrected sequences can have qualitatively different stellar responses.

The first case, denoted by case 1 in Fig.~\ref{fig:lnn_maxwell_mr}, represents a less extreme Maxwell-corrected sequence. Although the Maxwell construction introduces a coexistence plateau, the corresponding mass--radius curve does not immediately terminate at the transition point. The mass still increases slightly after the transition, indicating that a finite, although limited, hyperonic branch remains after the construction.

The second case, denoted by case 2, is qualitatively different. This case is obtained by modifying the $\Lambda NN$ density-dependent parameters to $(A_3,\beta)=(800,0.25)$. After the Maxwell construction, the mass--radius curve bends downward almost immediately after the onset or coexistence point. Consequently, the maximum mass is effectively determined by the onset point itself. In this case, the $\Lambda NN$ modification changes where the hyperonic transition occurs, but it does not generate an extended stable $\Lambda$-rich core that supports a larger mass.

This behavior has an important consequence for interpreting the mass--radius curves. For onset-controlled Maxwell-corrected sequences such as case 2, the stellar track follows the pure nucleonic or weakly hyperonic branch up to the transition point and then rapidly turns downward. Different $\Lambda NN$ parameter choices can therefore give different onset masses, while the observable part of the mass--radius track before the transition remains nearly degenerate with the no-$\Lambda$ sequence. This degeneracy should be kept in mind when interpreting both the deterministic maximum-mass map and any later statistical analysis based on global mass--radius behavior.

A similar, but distinct, issue appears in one-extremum branch-limited cases. Here the EOS does not provide two monotonic branches to be connected by a Maxwell construction. Instead, the neutron-star sequence follows the no-$\Lambda$ branch almost up to the $\Lambda$-onset point and then reaches the branch limit after sampling only a very short hyperonic extension. Operationally, such configurations can be regarded as stars with a very small $\Lambda$-rich inner core. Although they are not Maxwell-construction candidates by themselves, they produce a similar near-degeneracy of the mass--radius curves, because the observable part of the sequence remains close to the no-$\Lambda$ curve until the onset point.

The sharp bending in the representative Maxwell-constructed sequences reflects the replacement of the nonmonotonic segment by a constant-pressure coexistence segment. This example is used to show how a two-extrema branch can alter the corresponding mass--radius response once a Maxwell construction is imposed. It does not replace the full $(\beta,A_3)$-plane scan by a Maxwell-corrected map.

We now turn to the maximum-mass panels, Figs.~\ref{fig:lnn_mmax_extremum_count}(a) and \ref{fig:lnn_mmax_extremum_count}(d). In the mechanically admissible regions with $N_{\rm ext}^{\rm NS}=0$, identified from Figs.~\ref{fig:lnn_mmax_extremum_count}(b) and \ref{fig:lnn_mmax_extremum_count}(e), the $M_{\max}$ contours reflect the combined effect of the shifted $\Lambda$-onset density and the post-onset pressure response. Increasing $A_3$ generally enhances the density-dependent repulsion in the $N\Lambda$ sector, but because the same term also changes the onset condition, the resulting mass response is not simply equivalent to a smooth high-density stiffening. The contour pattern must therefore be read together with the extremum-count maps.

For $S\Lambda\Lambda1'$, the branch-limited and Maxwell-candidate regions identified in Figs.~\ref{fig:lnn_mmax_extremum_count}(b) and \ref{fig:lnn_mmax_extremum_count}(c) occupy a significant part of the mass map in Fig.~\ref{fig:lnn_mmax_extremum_count}(a) and are especially important for its interpretation. In these regions, the $M_{\max}$ contours are partly controlled by the onset, branch termination, or coexistence point, and not only by the stiffness of an extended stable hyperonic core. Therefore, the apparent recovery of $M_{\max}$ in this part of the scan should not be interpreted as a purely smooth stiffening mechanism. Instead, it indicates that the $\Lambda NN$ term can push the EOS into a regime where branch-structure effects become important for the interpretation of the mass map.

For $S\Lambda\Lambda3'$, the $M_{\max}$ contours in Fig.~\ref{fig:lnn_mmax_extremum_count}(d) are more directly interpretable over most of the parameter plane, because $N_{\rm ext}^{\rm NS}=0$ in a much larger region, as shown in Fig.~\ref{fig:lnn_mmax_extremum_count}(e). In this case, the $\Lambda NN$ term can increase the maximum mass primarily through the onset shift and the post-onset pressure response without immediately limiting the neutron-star branch. Nevertheless, the small nonmonotonic region near the edge of the scan in Figs.~\ref{fig:lnn_mmax_extremum_count}(e) and \ref{fig:lnn_mmax_extremum_count}(f) shows that even this reference interaction is not completely free from branch-structure effects.

The $2M_\odot$ contour in the maximum-mass panels, Figs.~\ref{fig:lnn_mmax_extremum_count}(a) and \ref{fig:lnn_mmax_extremum_count}(d), is again used only as a visual benchmark for mass recovery. In the $\Lambda NN$ scan, however, its interpretation is more delicate than in the $\Lambda\Lambda N$ scan. If the contour lies in a region with $N_{\rm ext}^{\rm NS}=0$, identified from Figs.~\ref{fig:lnn_mmax_extremum_count}(b) and \ref{fig:lnn_mmax_extremum_count}(e), it indicates ordinary mass recovery along a mechanically admissible branch. If it lies in a region with $N_{\rm ext}^{\rm NS}=1$ or in a Maxwell-candidate region, the corresponding mass is affected by branch termination or by the candidate coexistence structure. Thus, the $2M_\odot$ contour should not be read as a unique allowed boundary in the $(\beta,A_3)$ plane.

The present $\Lambda NN$ scan therefore shows a qualitatively different response from the $\Lambda\Lambda N$ scan. Because the $\Lambda NN$ density-dependent term is onset-active, it modifies the $\Lambda$-onset density and the post-onset EOS simultaneously. This produces organized regions of branch limitation and Maxwell-construction candidates, especially for the $S\Lambda\Lambda1'$ reference interaction. Consequently, the maximum-mass response in the $\Lambda NN$ scan cannot be interpreted from the $M_{\max}$ map alone. It must be read together with the onset shift, the extremum-count maps, and the possible phase-construction treatment.

\section{Bayesian and machine-learning analysis of hyperonic interaction parameters}
\label{sec:bayesian_ml}

In this section, we present the exploratory Bayesian posterior distributions and the machine-learning-based sensitivity analysis of the hyperonic EDF parameters. The Bayesian analysis is based only on neutron-star mass--radius information and is therefore not intended as a precision determination of the hyperonic interaction parameters. Instead, the purpose is to identify broad parameter tendencies that are compatible with the adopted mass--radius likelihood and to diagnose which hyperonic parameters most strongly affect selected neutron-star observables.

\subsection{Bayesian inference and machine-learning setup}
\label{subsec:bayesian_ml_methods}

Bayesian inference provides a systematic framework for connecting neutron-star mass--radius information to the dense-matter equation of state (EOS). Such analyses have been widely used in neutron-star EOS studies, including approaches that construct likelihoods from observational posterior samples in the mass--radius plane \cite{riley2018parametrized,raaijmakers2025neost}. Bayesian methods have also been applied to hyperonic neutron-star matter. For example, Bayesian analyses of hyperonic relativistic mean-field models have been used to investigate possible signatures of hyperons in neutron-star interiors \cite{malik2022hyperonBayes}. In the Skyrme framework, recent work has performed a Bayesian analysis of $\Lambda\Lambda$ and $\Lambda\Lambda N$ interaction parameters using hypernuclear and astrophysical constraints \cite{sun2026hyperonSkyrmeBayes}. A closely related study has also constrained the $\Lambda\Lambda$ interaction in a KIDS-based Skyrme energy-density-functional framework by combining terrestrial double-$\Lambda$ hypernuclear information with neutron-star mass--radius constraints \cite{tanimura2026lambdalambda}.

The Bayesian analysis in the present work is exploratory in scope. We use neutron-star mass--radius information alone to examine how the hyperonic Skyrme EDF parameters are mapped into the astrophysically relevant region for two nucleonic baselines, SLy230b and BSk24. We do not use the Bayesian evidence for model comparison. Instead, the posterior samples are used to identify broad parameter tendencies and the sensitivity of selected neutron-star observables to the $N\Lambda$ and $\Lambda\Lambda$ sectors. A more quantitative inference that combines mass--radius information with hypernuclear and nuclear data is left for future work.

For a model $\mathcal{M}$, the posterior distribution of the sampled hyperonic EDF parameters is defined up to normalization as
\begin{align}
p(\boldsymbol{\theta}\mid\mathcal{D},\mathcal{M}) \propto \mathcal{L}(\mathcal{D}\mid\boldsymbol{\theta},\mathcal{M})\pi(\boldsymbol{\theta}\mid\mathcal{M}),
\label{eq:bayes_posterior_mr}
\end{align}
where $\boldsymbol{\theta}$ is the parameter vector, $\mathcal{D}$ denotes the observational mass--radius information, $\pi$ is the prior, and $\mathcal{L}$ is the likelihood. The posterior was explored with \texttt{dynesty}, a public nested-sampling package \cite{speagle2020dynesty}, which is used here as a practical posterior-exploration method that does not assume a Gaussian posterior distribution.

We considered two inference setups. The first setup is a seven-dimensional $N\Lambda$-sector scan, denoted by ``7dim ($\Lambda\Lambda$ fixed)'' in the figures. In this setup, the two-body $\Lambda\Lambda$ sector is fixed to the S$\Lambda\Lambda$1$^{\prime}$ reference choice, while the three two-body $N\Lambda$ EDF coefficients $A_0$, $A_1$, and $A_2$ are varied together with the four hyperonic three-body parameters. These four common parameters are $(A_3,\beta)$ for the effective $\Lambda NN$ term and $(C_3,\gamma)$ for the effective $\Lambda\Lambda N$ term. The sampled parameter vector is
\begin{align}
\boldsymbol{\theta}_{7{\rm D}}=(\underbrace{A_0,A_1,A_2}_{N\Lambda},\underbrace{A_3,\beta}_{\Lambda NN},\underbrace{C_3,\gamma}_{\Lambda\Lambda N}).
\label{eq:theta_7d_scan}
\end{align}
The pair $(A_3,\beta)$ controls the effective $\Lambda NN$ contribution, while $(C_3,\gamma)$ controls the effective $\Lambda\Lambda N$ contribution. This setup is seven-dimensional because the three two-body $N\Lambda$ coefficients are sampled together with the four common hyperonic three-body parameters.

The second setup is a six-dimensional $\Lambda\Lambda$-sector scan, denoted by ``6dim ($N\Lambda$ fixed)'' in the figures. In this setup, the two-body $N\Lambda$ sector is fixed to the SLL4$^{\prime}$ reference choice, while the two-body $\Lambda\Lambda$ coefficients $C_0$ and $C_1$ are varied together with the same four hyperonic three-body parameters, namely $(A_3,\beta)$ for the $\Lambda NN$ term and $(C_3,\gamma)$ for the $\Lambda\Lambda N$ term. The sampled parameter vector is
\begin{align}
\boldsymbol{\theta}_{6{\rm D}}=(\underbrace{C_0,C_1}_{\Lambda\Lambda},\underbrace{A_3,\beta}_{\Lambda NN},\underbrace{C_3,\gamma}_{\Lambda\Lambda N}).
\label{eq:theta_6d_scan}
\end{align}
Here $C_0$ and $C_1$ specify the two-body $\Lambda\Lambda$ sector in the EDF-coefficient convention. The remaining $\Lambda\Lambda$ coefficient $C_2$ was fixed to zero in the present scan. This setup is six-dimensional because the two two-body $\Lambda\Lambda$ coefficients are sampled together with the same four hyperonic three-body parameters used in the seven-dimensional setup.

The scan design can be summarized as
\begin{align}
\mathcal{S}_{7{\rm D}}&=\{{\rm SLy230b},{\rm BSk24}\}\times\boldsymbol{\theta}_{7{\rm D}},
\label{eq:scan_design_7d}\\
\mathcal{S}_{6{\rm D}}&=\{{\rm SLy230b},{\rm BSk24}\}\times\boldsymbol{\theta}_{6{\rm D}}.
\label{eq:scan_design_6d}
\end{align}
In $\mathcal{S}_{7{\rm D}}$, the two-body $\Lambda\Lambda$ sector is fixed to S$\Lambda\Lambda$1$^{\prime}$. In $\mathcal{S}_{6{\rm D}}$, the two-body $N\Lambda$ sector is fixed to SLL4$^{\prime}$. Thus, the four-dimensional three-body sector, $(A_3,\beta,C_3,\gamma)$, is common to both setups. The difference between the two setups is which two-body hyperonic sector is varied: the seven-dimensional scan varies the $N\Lambda$ two-body sector, whereas the six-dimensional scan varies the $\Lambda\Lambda$ two-body sector. This notation is used consistently below in the posterior, mass--radius, and SHAP figures.

Independent uniform priors were assigned to the sampled parameters. In the seven-dimensional scan, $A_0$, $A_1$, and $A_2$ are varied together with the four hyperonic three-body parameters. In the six-dimensional scan, $C_0$ and $C_1$ are varied with the same four hyperonic three-body parameters, while the two-body $N\Lambda$ sector is fixed to the SLL4$^{\prime}$ reference. The prior ranges are summarized in Table~\ref{tab:bayesian_prior_ranges}.

\begin{table}[tbp]
\centering
\caption{Prior ranges used in the exploratory Bayesian inference. Here 7D denotes the seven-dimensional scan with the two-body $\Lambda\Lambda$ sector fixed to S$\Lambda\Lambda$1$^{\prime}$, and 6D denotes the six-dimensional scan with the two-body $N\Lambda$ sector fixed to SLL4$^{\prime}$. The four parameters common to both setups are $(A_3,\beta,C_3,\gamma)$, where $(A_3,\beta)$ controls the $\Lambda NN$ term and $(C_3,\gamma)$ controls the $\Lambda\Lambda N$ term.}
\label{tab:bayesian_prior_ranges}
\begin{tabular}{c c c}
\hline\hline
Setup & Parameter & Prior range \\
\hline
7D & $A_0$ & $[-550,\,-200]$ \\
7D & $A_1$ & $[-5,\,55]$ \\
7D & $A_2$ & $[0,\,80]$ \\
Both & $A_3$ & $[0,\,700]$ \\
Both & $\beta$ & $[0,\,1]$ \\
Both & $C_3$ & $[0,\,700]$ \\
Both & $\gamma$ & $[0,\,1]$ \\
6D & $C_0$ & $[-250,\,-10^{-6}]$ \\
6D & $C_1$ & $[10^{-6},\,140]$ \\
\hline\hline
\end{tabular}
\end{table}

For each sampled parameter vector, the beta-equilibrated EOS was constructed using the EDF described above, and the corresponding mass--radius relation was obtained by solving the TOV equations. Samples for which the EOS construction or TOV integration failed were assigned a negligible likelihood. No additional hard lower cut on the maximum mass was imposed. Instead, compatibility with neutron-star observations was evaluated directly through the mass--radius likelihood described below.

The observational information was represented by approximate effective likelihoods in the $(M,R)$ plane. Specifically, we constructed two-dimensional KDEs from the reported mass--radius posterior samples. This use of posterior samples as likelihood summaries follows the practical KDE-based workflow adopted in EOS-inference frameworks such as NEoST \cite{raaijmakers2025neost}, while the associated prior dependence of such posterior-based constructions has been discussed in Ref.~\cite{riley2018parametrized}. The procedure should therefore be interpreted as an approximation: the reported posterior density is proportional to the observational likelihood multiplied by the prior adopted in the original pulse-profile analysis. A prior correction would be required for a fully prior-independent likelihood reconstruction. In the present exploratory study, we use the KDEs of the reported posterior samples as effective likelihood summaries and do not apply an additional prior correction. We used mass--radius posterior samples from the NICER pulse-profile analyses of PSR~J0030+0451, PSR~J0740+6620, and PSR~J0437--4715 \cite{riley2019j0030,riley2021j0740,choudhury2024j0437}. For each source $i$, the posterior samples were converted into a kernel density estimate $p_i(M,R)$. When weighted posterior samples were provided, equal-weight samples were generated by systematic resampling before constructing the KDE.

To reduce sensitivity to far posterior tails, the likelihood was evaluated over the central $98\%$ mass interval of each observational posterior,
\begin{align}
I_i=\left[M_{i,1\%},M_{i,99\%}\right],\qquad \Delta M_i=M_{i,99\%}-M_{i,1\%}.
\label{eq:obs_mass_interval}
\end{align}
For a calculated EOS that gives a single-valued relation $R_{\boldsymbol{\theta}}(M)$ over the relevant mass interval, the likelihood contribution from source $i$ was evaluated as
\begin{align}
\mathcal{L}_i(\boldsymbol{\theta})=\frac{1}{\Delta M_i}\int_{I_i\cap I_{\boldsymbol{\theta}}}p_i\!\left(M,R_{\boldsymbol{\theta}}(M)\right)\,dM,
\label{eq:single_branch_likelihood}
\end{align}
where $I_{\boldsymbol{\theta}}$ is the mass interval covered by the calculated mass--radius curve. The normalization was kept fixed to the full observational interval $\Delta M_i$; consequently, EOSs covering only part of the observational mass support were penalized rather than renormalized over the overlapping interval alone.

Some calculated mass--radius curves are not single-valued in $M$ when the sequence is followed in order of increasing central pressure. In such cases, we split the curve into monotonic-in-$M$ segments and write the radius on segment $b$ as $R_{\boldsymbol{\theta},b}(M)$. This segmentation is an operational prescription for evaluating a multi-valued $R(M)$ relation in the likelihood. It should not be interpreted as a claim that every segment, including a post-maximum-mass descending segment, corresponds to a distinct stable stellar family.

For a multi-valued mass--radius curve, the segment label was treated as an unobserved discrete variable and marginalized over at fixed mass. The source likelihood was then defined as
\begin{align}
\mathcal{L}_i(\boldsymbol{\theta})=\frac{1}{\Delta M_i}\int_{I_i}\left[\frac{1}{N_{\boldsymbol{\theta}}(M)}\sum_{b\in\mathcal{B}_{\boldsymbol{\theta}}(M)}p_i\!\left(M,R_{\boldsymbol{\theta},b}(M)\right)\right]dM.
\label{eq:branch_marginalized_likelihood}
\end{align}
Here $\mathcal{B}_{\boldsymbol{\theta}}(M)$ is the set of monotonic-in-$M$ segments available at mass $M$, $N_{\boldsymbol{\theta}}(M)$ is the number of such segments, and the integrand is set to zero at masses where no segment exists. The fixed normalization by $\Delta M_i$ again penalizes missing mass support. This prescription corresponds to assigning equal prior weight to the available segments at fixed mass and marginalizing over the unobserved segment label. Equation~\eqref{eq:single_branch_likelihood} is recovered as the single-branch limit of Eq.~\eqref{eq:branch_marginalized_likelihood}.

The total likelihood was obtained by multiplying the individual source likelihoods,
\begin{align}
\mathcal{L}(\mathcal{D}\mid\boldsymbol{\theta},\mathcal{M})=\prod_i\mathcal{L}_i(\boldsymbol{\theta}),\qquad \ln\mathcal{L}=\sum_i\ln\mathcal{L}_i.
\label{eq:total_mr_likelihood}
\end{align}
The posterior samples obtained from this likelihood were then used both for the one-dimensional marginalized posterior distributions and for the weighted posterior mass--radius curves shown below.

To summarize the nonlinear dependence of neutron-star observables on the hyperonic EDF parameters, we also trained XGBoost regressors on the posterior samples \cite{chen2016xgboost}. The target observables were the maximum gravitational mass $M_{\max}$ and the radius of a $2.0M_\odot$ neutron star, $R_{2.0}$. We then computed SHAP values for the trained surrogate models \cite{lundberg2017shap}. The SHAP analysis is used only as a model-interpretability diagnostic for the trained regressors. It should not be interpreted as an independent proof of parameter causality, but it provides a compact way to identify which parameters control the surrogate predictions within the posterior region.

The results are organized as follows. We first discuss the marginalized posterior distributions of the hyperonic EDF parameters, then examine the posterior mass--radius relations, and finally use XGBoost--SHAP surrogate analysis to summarize the parameter sensitivity of $M_{\max}$ and $R_{2.0}$.

\subsection{Posterior distributions of the hyperonic EDF parameters}

\begin{figure*}[tbp]
\centering
\includegraphics[width=\textwidth]{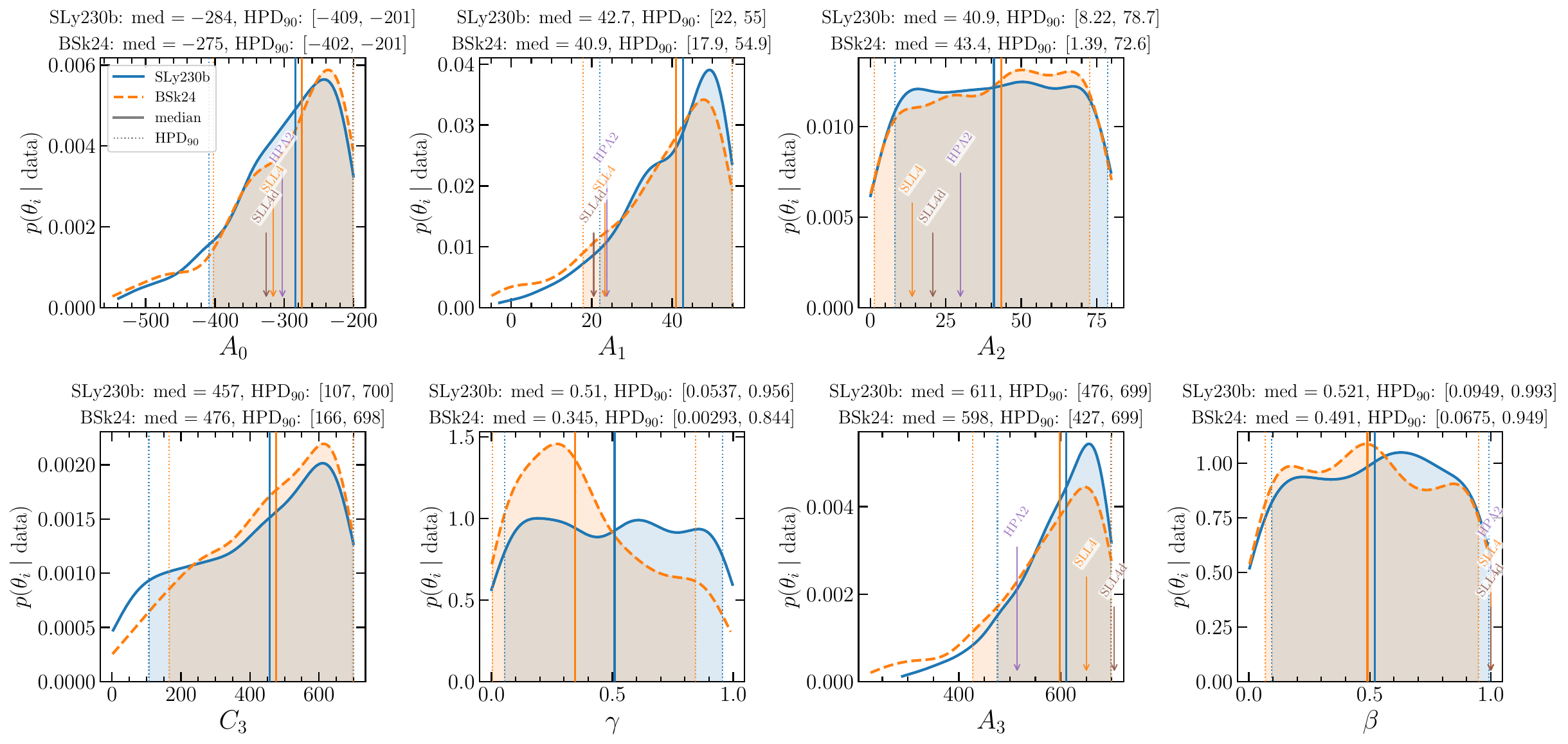}
\caption{Marginalized one-dimensional posterior distributions for the seven-dimensional $N\Lambda$-sector scan. In this setup, the two-body $\Lambda\Lambda$ sector is fixed to the S$\Lambda\Lambda$1$^{\prime}$ reference choice, while $(A_0,A_1,A_2,A_3,\beta,C_3,\gamma)$ are varied. The four parameters $(A_3,\beta,C_3,\gamma)$ are the common hyperonic three-body sector, where $(A_3,\beta)$ controls the $\Lambda NN$ term and $(C_3,\gamma)$ controls the $\Lambda\Lambda N$ term. The solid and dashed curves correspond to the SLy230b and BSk24 nucleonic baselines, respectively. Vertical solid lines indicate the posterior medians, and dotted lines indicate the $90\%$ highest-posterior-density intervals. Reference $N\Lambda$ parameter sets are shown by colored vertical markers. The panel label SLL4d denotes the SLL4$^{\prime}$ parameter set.}
\label{fig:posteriors_7d_NLambda_scan}
\end{figure*}

Figure~\ref{fig:posteriors_7d_NLambda_scan} shows the marginalized one-dimensional posterior distributions for the seven-dimensional $N\Lambda$-sector scan. In this setup, the two-body $\Lambda\Lambda$ sector is fixed to the S$\Lambda\Lambda$1$^{\prime}$ reference choice, while the two-body $N\Lambda$ parameters $A_0$, $A_1$, and $A_2$ are varied together with the four hyperonic three-body parameters $(A_3,\beta,C_3,\gamma)$. The pair $(A_3,\beta)$ represents the $\Lambda NN$ term, and $(C_3,\gamma)$ represents the $\Lambda\Lambda N$ term. The two nucleonic baselines, SLy230b and BSk24, lead to broadly similar posterior shapes. The posterior of $A_0$ remains concentrated in the attractive region and has medians of approximately $-284$ and $-275$ for SLy230b and BSk24, respectively. The posterior of $A_1$ is shifted toward relatively large positive values, with medians around $42.7$ and $40.9$. By contrast, $A_2$ remains comparatively broad, indicating that the present mass--radius likelihood alone does not isolate a narrow preferred interval for this coefficient.

The three-body parameters show a clearer qualitative tendency. The strength $A_3$ of the effective $\Lambda NN$ contribution is strongly shifted toward the upper part of the prior range in both nucleonic baselines, with medians of approximately $611$ for SLy230b and $598$ for BSk24. The $\Lambda\Lambda N$ strength $C_3$ is also biased toward moderately large values, with medians of approximately $457$ and $476$. In contrast, the exponents $\beta$ and $\gamma$ remain broad. This suggests that the mass--radius likelihood mainly favors additional repulsion through the three-body strengths, while the corresponding exponents are only weakly constrained in the present exploratory setup.

\begin{figure*}[tbp]
\centering
\includegraphics[width=\textwidth]{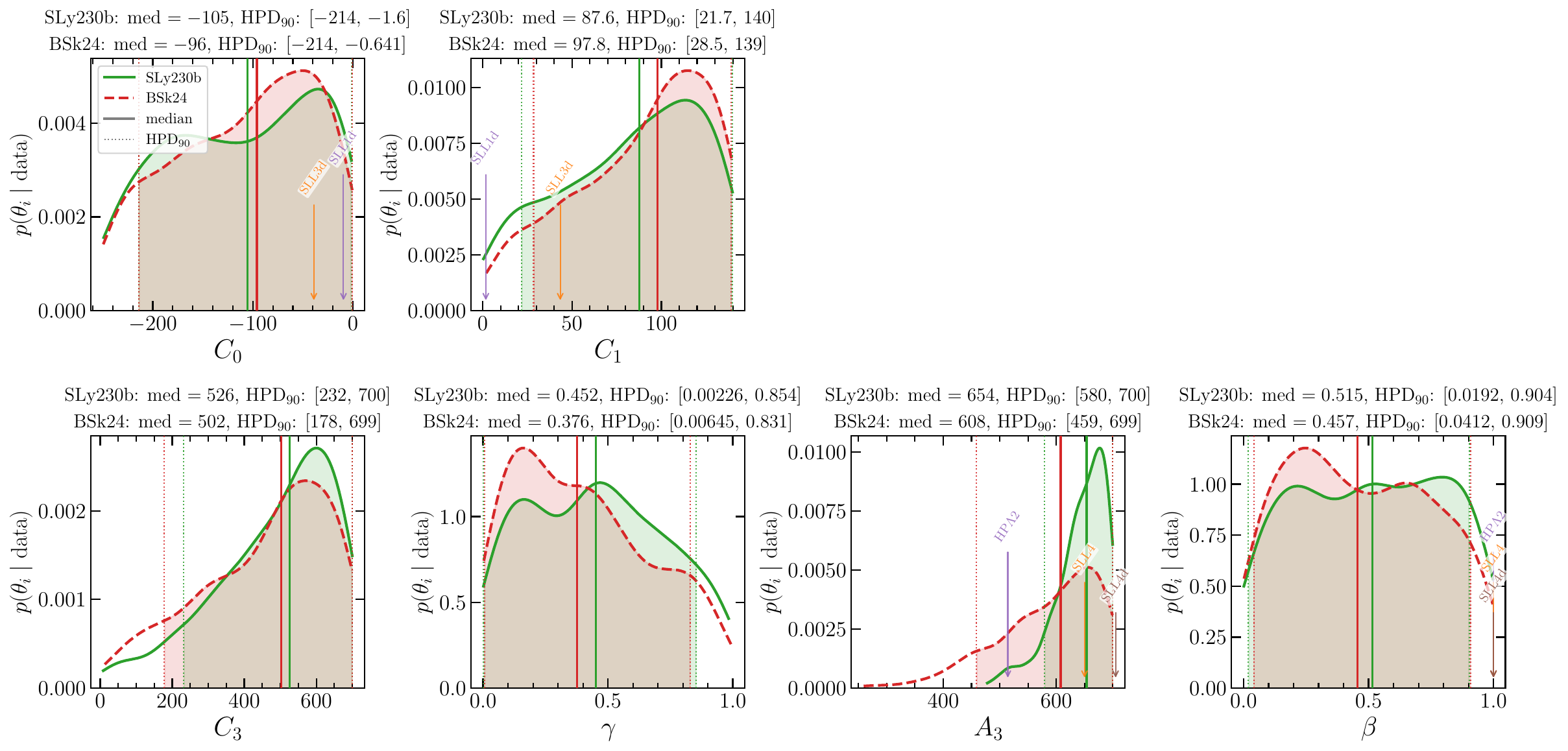}
\caption{Marginalized one-dimensional posterior distributions for the six-dimensional $\Lambda\Lambda$-sector scan. In this setup, the two-body $N\Lambda$ sector is fixed to the SLL4$^{\prime}$ reference choice, while $(C_0,C_1,A_3,\beta,C_3,\gamma)$ are varied. The four parameters $(A_3,\beta,C_3,\gamma)$ are the same common hyperonic three-body sector as in Fig.~\ref{fig:posteriors_7d_NLambda_scan}, where $(A_3,\beta)$ controls the $\Lambda NN$ term and $(C_3,\gamma)$ controls the $\Lambda\Lambda N$ term. The solid and dashed curves correspond to the SLy230b and BSk24 nucleonic baselines, respectively. Vertical solid lines indicate posterior medians, and dotted lines indicate the $90\%$ highest-posterior-density intervals. Reference $\Lambda\Lambda$ parameter sets are shown by colored vertical markers for the two-body $\Lambda\Lambda$ coefficients, while reference $N\Lambda$ parameter sets are shown where applicable for the $\Lambda NN$ parameters. The panel labels SLL1d and SLL3d denote S$\Lambda\Lambda$1$^{\prime}$ and S$\Lambda\Lambda$3$^{\prime}$, respectively, while SLL4d denotes SLL4$^{\prime}$.}
\label{fig:posteriors_6d_LLambda_scan}
\end{figure*}

Figure~\ref{fig:posteriors_6d_LLambda_scan} shows the corresponding marginalized posteriors for the six-dimensional $\Lambda\Lambda$-sector scan. In this setup, the two-body $N\Lambda$ sector is fixed to the SLL4$^{\prime}$ reference choice, while $C_0$ and $C_1$ are varied together with the same four hyperonic three-body parameters $(A_3,\beta,C_3,\gamma)$. Again, $(A_3,\beta)$ and $(C_3,\gamma)$ represent the $\Lambda NN$ and $\Lambda\Lambda N$ terms, respectively. The posterior of $C_0$ remains in the attractive region, with medians of approximately $-105$ for SLy230b and $-96$ for BSk24. The coefficient $C_1$ is shifted toward relatively large positive values, with medians of approximately $87.6$ and $97.8$. These tendencies indicate that, within the adopted prior ranges and likelihood, the $\Lambda\Lambda$ two-body sector is not driven toward a single narrow reference interaction but remains broadly distributed.

The three-body sector in Fig.~\ref{fig:posteriors_6d_LLambda_scan} shows the same overall pattern as in Fig.~\ref{fig:posteriors_7d_NLambda_scan}. The posterior distribution of $A_3$ is concentrated near large values, with medians of approximately $654$ for SLy230b and $608$ for BSk24. The posterior of $C_3$ also favors relatively large values, with medians around $526$ and $502$. The exponents $\beta$ and $\gamma$ remain broad. Therefore, across both scan types, the most robust qualitative feature is the preference for sizable three-body repulsion, rather than a sharply determined exponent.

\subsection{Posterior mass--radius relations}

\begin{figure}[tbp]
\centering
\includegraphics[width=\columnwidth]{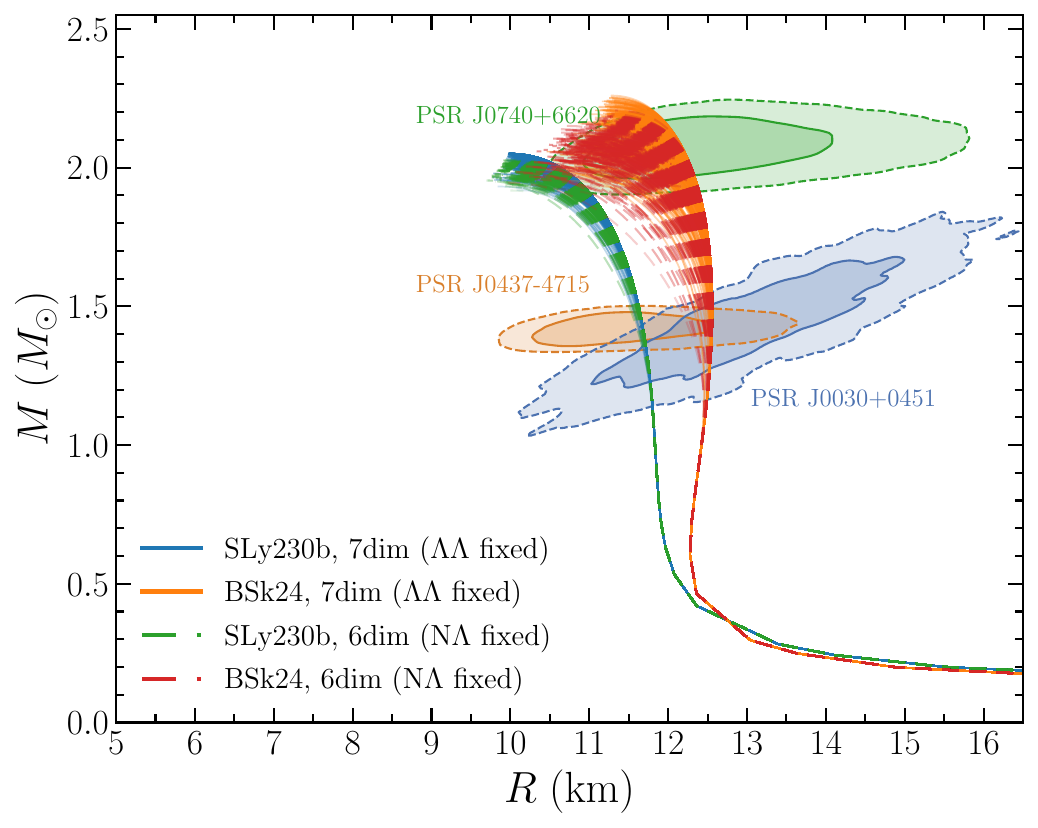}
\caption{Weighted posterior mass--radius curves for the four exploratory Bayesian cases. The shaded contours show the mass--radius posterior regions of PSR~J0030+0451, PSR~J0740+6620, and PSR~J0437--4715 used in the likelihood construction. The labels ``7dim ($\Lambda\Lambda$ fixed)'' denote the seven-dimensional scans in which $(A_0,A_1,A_2,A_3,\beta,C_3,\gamma)$ are varied with the two-body $\Lambda\Lambda$ sector fixed to S$\Lambda\Lambda$1$^{\prime}$. The labels ``6dim ($N\Lambda$ fixed)'' denote the six-dimensional scans in which $(C_0,C_1,A_3,\beta,C_3,\gamma)$ are varied with the two-body $N\Lambda$ sector fixed to SLL4$^{\prime}$. In both cases, the common four-dimensional three-body sector is $(A_3,\beta,C_3,\gamma)$, consisting of two parameters for the $\Lambda NN$ term and two parameters for the $\Lambda\Lambda N$ term.}
\label{fig:mr_posterior_weighted}
\end{figure}

Figure~\ref{fig:mr_posterior_weighted} shows the weighted posterior mass--radius curves for the four exploratory Bayesian cases, together with the NICER posterior contours used in the mass--radius likelihood. The four cases are the SLy230b and BSk24 versions of the seven-dimensional scan with the two-body $\Lambda\Lambda$ sector fixed to S$\Lambda\Lambda$1$^{\prime}$, and the SLy230b and BSk24 versions of the six-dimensional scan with the two-body $N\Lambda$ sector fixed to SLL4$^{\prime}$. In both scan types, the common four-dimensional three-body sector $(A_3,\beta,C_3,\gamma)$ is varied. The posterior curves occupy a relatively compact region around radii of roughly $11$--$12.5~{\rm km}$ for canonical and high-mass neutron stars. All four cases pass through the mass range relevant to PSR~J0740+6620 and PSR~J0437--4715, while the broader PSR~J0030+0451 posterior allows a wider radius region.

The similarity among the four posterior bands indicates that the adopted mass--radius information does not strongly distinguish the two scan types or the two nucleonic baselines at the level of this exploratory analysis. This is consistent with the broad one-dimensional posteriors shown in Figs.~\ref{fig:posteriors_7d_NLambda_scan} and \ref{fig:posteriors_6d_LLambda_scan}. The posterior mass--radius relations are mainly selected by the requirement that the hyperonic EOS remain sufficiently stiff to support massive neutron stars while retaining radii compatible with the NICER posterior regions. This explains why the three-body strengths, especially $A_3$ and $C_3$, tend to be shifted toward large values. In particular, when a large $C_3$ is required to keep the sequence compatible with the $2.0M_\odot$ scale, the resulting mass--radius curve can remain close to the corresponding hyperon-free nucleonic baseline over a substantial part of the sequence before the finite-$\Lambda$ contribution becomes dominant. This behavior implies that the nucleonic EDF baseline still plays an important role in shaping the posterior mass--radius relation, even when the inference mainly favors strong hyperonic three-body repulsion.

\subsection{XGBoost--SHAP sensitivity analysis}

\begin{figure}[tbp]
\centering
\includegraphics[width=\columnwidth]{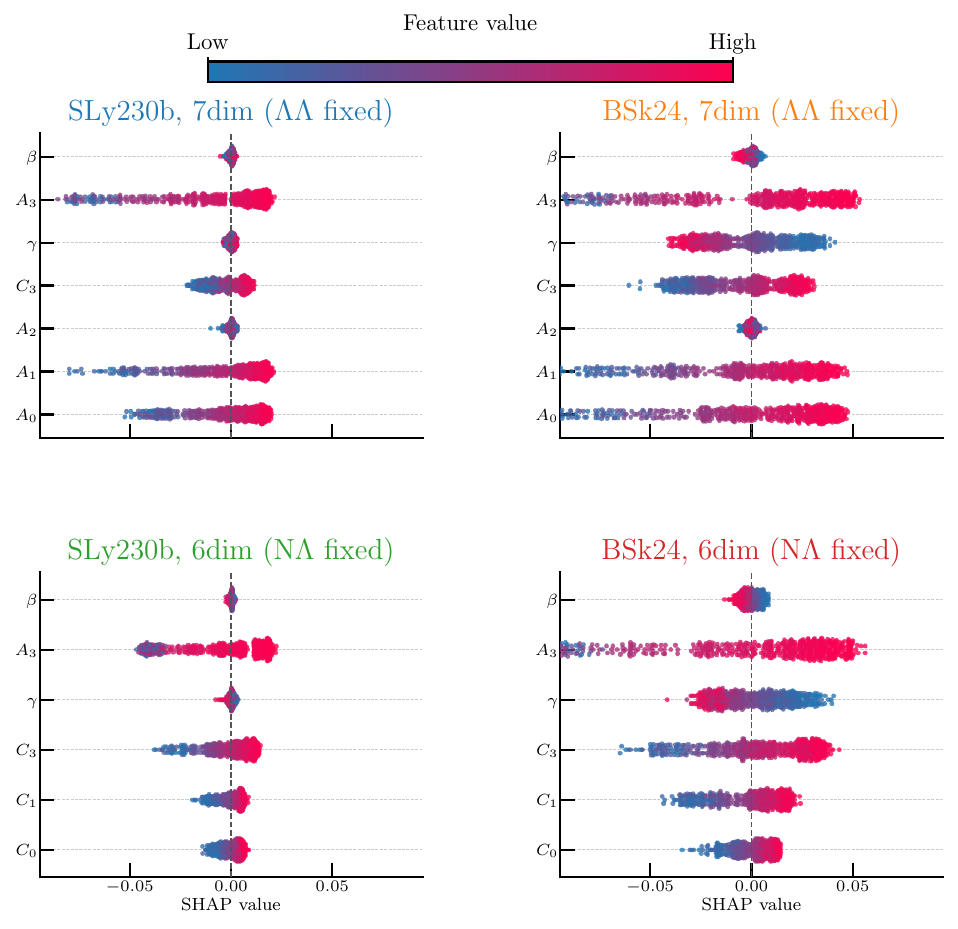}
\caption{SHAP beeswarm plots for the XGBoost surrogate model trained to predict $M_{\max}$. The upper panels correspond to the seven-dimensional scans in which $(A_0,A_1,A_2,A_3,\beta,C_3,\gamma)$ are varied with the two-body $\Lambda\Lambda$ sector fixed to S$\Lambda\Lambda$1$^{\prime}$. The lower panels correspond to the six-dimensional scans in which $(C_0,C_1,A_3,\beta,C_3,\gamma)$ are varied with the two-body $N\Lambda$ sector fixed to SLL4$^{\prime}$. The four parameters $(A_3,\beta,C_3,\gamma)$ are common to all panels and represent the two hyperonic three-body terms: $(A_3,\beta)$ for $\Lambda NN$ and $(C_3,\gamma)$ for $\Lambda\Lambda N$. The horizontal axis shows the SHAP value, and the color indicates the value of the corresponding input feature. Parameters with wider SHAP distributions have a larger influence on the surrogate prediction within the sampled posterior region.}
\label{fig:shap_Mmax}
\end{figure}

To further diagnose the parameter dependence of the posterior samples, we trained XGBoost regressors for two representative observables, $M_{\max}$ and $R_{2.0}$, and computed SHAP values for the trained surrogate models. Figure~\ref{fig:shap_Mmax} shows the SHAP beeswarm plots for $M_{\max}$. The four panels correspond to the same four Bayesian cases as in Fig.~\ref{fig:mr_posterior_weighted}: the two seven-dimensional scans with the two-body $\Lambda\Lambda$ sector fixed to S$\Lambda\Lambda$1$^{\prime}$ and the two six-dimensional scans with the two-body $N\Lambda$ sector fixed to SLL4$^{\prime}$. The dominant features differ somewhat between scan types and nucleonic baselines, but a common trend is that the three-body strength parameters, particularly $A_3$ and $C_3$, have large SHAP spreads. This indicates that they are among the main parameters controlling the maximum mass within the posterior region.

In the seven-dimensional $N\Lambda$-sector scan, the two-body parameters $A_0$ and $A_1$ also contribute appreciably to the $M_{\max}$ prediction, especially for the BSk24 baseline. In the six-dimensional $\Lambda\Lambda$-sector scan, $C_0$ and $C_1$ show weaker but still visible contributions compared with the three-body strengths. The exponents $\beta$ and $\gamma$ generally produce narrower SHAP distributions than the corresponding strength parameters. This behavior is consistent with the posterior distributions: the likelihood is more sensitive to the overall amount of repulsion in the three-body sector than to the precise exponent controlling it.

\begin{figure}[tbp]
\centering
\includegraphics[width=\columnwidth]{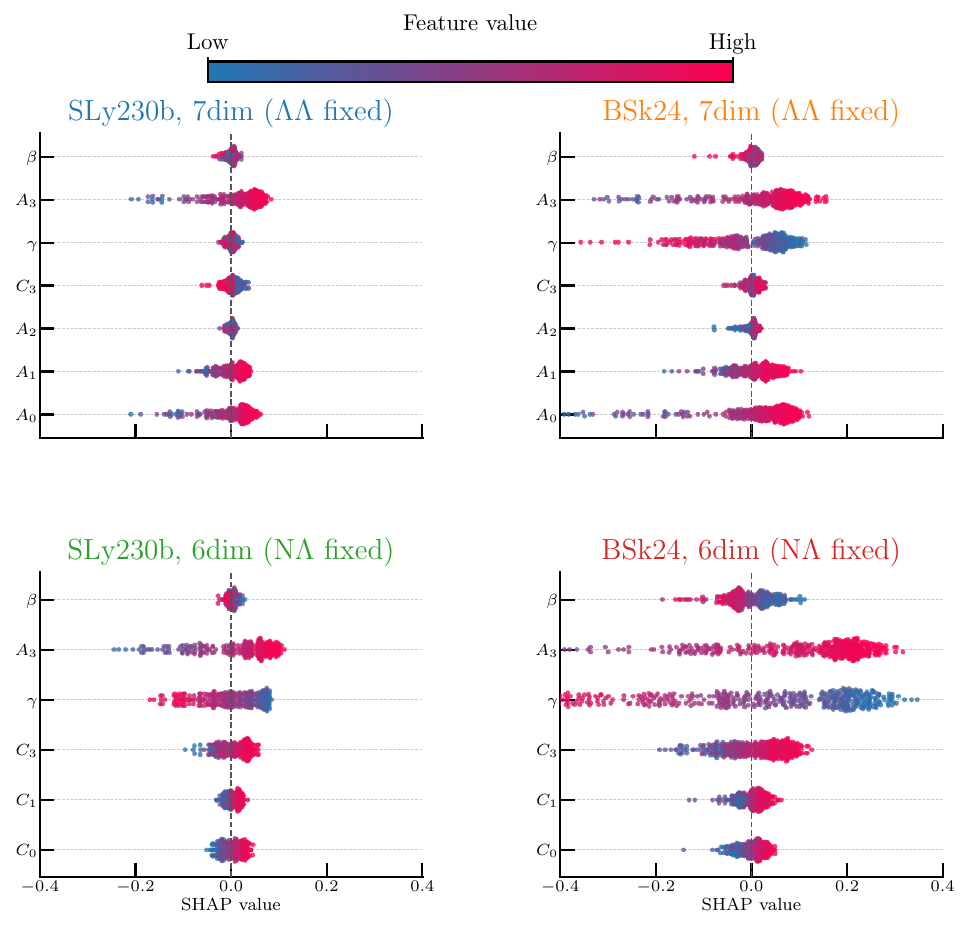}
\caption{SHAP beeswarm plots for the XGBoost surrogate model trained to predict $R_{2.0}$. The upper panels correspond to the seven-dimensional scans in which $(A_0,A_1,A_2,A_3,\beta,C_3,\gamma)$ are varied with the two-body $\Lambda\Lambda$ sector fixed to S$\Lambda\Lambda$1$^{\prime}$. The lower panels correspond to the six-dimensional scans in which $(C_0,C_1,A_3,\beta,C_3,\gamma)$ are varied with the two-body $N\Lambda$ sector fixed to SLL4$^{\prime}$. The four parameters $(A_3,\beta,C_3,\gamma)$ are common to all panels and represent the two hyperonic three-body terms: $(A_3,\beta)$ for $\Lambda NN$ and $(C_3,\gamma)$ for $\Lambda\Lambda N$. The horizontal axis shows the SHAP value, and the color indicates the value of the corresponding input feature. The larger SHAP spread compared with Fig.~\ref{fig:shap_Mmax} indicates that the radius at $2.0M_\odot$ is sensitive to multiple hyperonic parameters within the posterior region.}
\label{fig:shap_R20}
\end{figure}

Figure~\ref{fig:shap_R20} shows the SHAP beeswarm plots for $R_{2.0}$. Compared with $M_{\max}$, the SHAP values for $R_{2.0}$ are distributed over a much wider scale, indicating that the radius of a massive neutron star is more sensitive to variations among several hyperonic parameters within the posterior region. A common feature of all four panels is the importance of $A_3$: larger values of $A_3$ generally give positive SHAP contributions to $R_{2.0}$, whereas smaller values tend to reduce the predicted radius. This trend is consistent with the role of the $\Lambda NN$ three-body term in maintaining a stiffer high-mass branch.

The exponent $\gamma$ also becomes more visible in $R_{2.0}$ than in $M_{\max}$, especially for the BSk24 baseline and for the six-dimensional scan. In these cases, the SHAP distribution of $\gamma$ extends over a range comparable to, or even larger than, that of some strength parameters, showing that the radius at $2.0M_\odot$ is sensitive not only to the overall magnitude of the three-body repulsion but also to how the $\Lambda\Lambda N$ contribution changes along the stellar sequence. By contrast, $\beta$ remains relatively localized around zero in all panels, suggesting a weaker direct impact on $R_{2.0}$ within the present posterior samples.

The scanned two-body parameters also contribute in a setup-dependent way. In the seven-dimensional scans, the $N\Lambda$ coefficients $A_0$ and $A_1$ show appreciable SHAP spreads, particularly for the BSk24 baseline, indicating that the $N\Lambda$ two-body sector still affects the radius once the sequence reaches the high-mass region. In the six-dimensional scans, the $\Lambda\Lambda$ coefficients $C_0$ and $C_1$ have more modest SHAP spreads, while $C_3$ contributes more visibly, especially for BSk24. Overall, Fig.~\ref{fig:shap_R20} suggests that $R_{2.0}$ is controlled by a broader combination of parameters than $M_{\max}$, with $A_3$ remaining robustly important and $\gamma$ becoming particularly relevant for the radius response.

The SHAP results should be interpreted as an explanation of the trained surrogate models, not as an independent physical proof of parameter causality. Nevertheless, they provide a useful summary of the nonlinear response of the mass--radius observables to the hyperonic EDF parameters. In both $M_{\max}$ and $R_{2.0}$, the three-body terms are repeatedly identified as important features. This supports the qualitative interpretation drawn from the posterior distributions: in the present exploratory mass--radius-only analysis, the data primarily select EOSs with sufficient hyperonic repulsion, while leaving substantial degeneracy among the detailed hyperonic interaction parameters.

Overall, the Bayesian and SHAP analyses point to the same conclusion. The current mass--radius information alone can identify broad tendencies, such as the preference for sizable hyperonic three-body repulsion, but it does not provide a precise determination of all $N\Lambda$ and $\Lambda\Lambda$ parameters. A more restrictive inference will require combining the astrophysical likelihood with hypernuclear data and nuclear constraints, as well as using a sampling setup designed for quantitative evidence-based comparison.

\section{Summary and Outlook}
\label{sec:summary_outlook}

We have investigated the effects of hyperonic density-dependent terms on neutron-star equations of state within a Skyrme energy-density-functional framework. Cold beta-equilibrated matter including nucleons, leptons, and $\Lambda$ hyperons was constructed, and the resulting EOSs were used to solve the Tolman--Oppenheimer--Volkoff equations. The $\Lambda NN$ and $\Lambda\Lambda N$ density-dependent terms were varied separately, so that their effects on the $\Lambda$-onset behavior, the post-onset EOS, the mass--radius relation, and the maximum mass could be isolated.

A central element of the analysis was to classify the calculated EOS
curves in the $P$-$\varepsilon$ plane according to their
monotonicity and extremum structure. We distinguished monotonic branches, one-extremum branch-limited cases, and two-extrema Maxwell-construction candidates. The extremum-count maps $N_{\rm ext}^{\rm NS}$ and $N_{\rm ext}^{\rm full}$ were introduced to separate extrema encountered within the neutron-star-relevant pressure range from extrema that appear only in the full high-density EOS. These maps were used as diagnostics for interpreting the $M_{\max}$ maps, rather than as additional observables.

In the $\Lambda\Lambda N$ scan, the density-dependent term is onset-inactive in the local continuous onset condition because its contribution is of order $n_\Lambda^2$. Nevertheless, it modifies the finite-$\Lambda$ post-onset EOS.
In the updated $\Lambda\Lambda$-sector references considered here,
increasing $C_3$ generally enhances the post-onset pressure and raises
$M_{\max}$ in the mechanically admissible regions, whereas increasing
$\gamma$ tends to reduce this enhancement at fixed $C_3$.
Some extrema appear in the full calculated EOS, but many of them lie outside the pressure range relevant to the maximum-mass configuration, which makes the distinction between $N_{\rm ext}^{\rm full}$ and $N_{\rm ext}^{\rm NS}$ essential.

In the $\Lambda NN$ scan, the density-dependent term is onset-active and therefore changes the $\Lambda$-onset density and the post-onset EOS simultaneously. This produces a more intricate relation between the $M_{\max}$ map and the underlying EOS branch structure. For the $S\Lambda\Lambda1'$ reference interaction, branch-limited regions and Maxwell-construction candidates occupy a sizable part of the parameter plane, 
whereas the $S\Lambda\Lambda3'$ reference interaction remains
mechanically admissible over a wider region and therefore gives a more
directly interpretable mass response.
Representative Maxwell constructions illustrate how a two-extrema branch can alter the corresponding mass--radius response, but a Maxwell-corrected $M_{\max}$ map over the full $(\beta,A_3)$ plane was not constructed in this work.

The main physical conclusion is that mass recovery in hyperonic neutron stars is not a single mechanism. In mechanically admissible monotonic regions, an increase of $M_{\max}$ can be interpreted as ordinary high-density stiffening of the hyperonic EOS. In branch-limited or Maxwell-candidate regions, however, the same value of $M_{\max}$ or the same $2M_\odot$ contour can also reflect the $\Lambda$-onset point, the branch limit, or the candidate coexistence structure. The maximum-mass maps must therefore be read together with the onset behavior and the extremum-count maps.

The exploratory Bayesian and machine-learning analyses provide a complementary view of these deterministic results. Using neutron-star mass--radius information alone, the posterior samples favor sizable hyperonic three-body repulsion but retain broad degeneracies among the detailed $N\Lambda$ and $\Lambda\Lambda$ parameters. The posterior mass--radius relations are similar among the four scan cases, indicating that the adopted mass--radius information does not strongly distinguish the two scan types or the two nucleonic baselines at the level of the present exploratory analysis. The XGBoost--SHAP diagnostics identify the three-body strength parameters, especially $A_3$ and $C_3$, as important controls of the surrogate predictions, while $R_{2.0}$ is sensitive to a broader combination of parameters than $M_{\max}$. These results support the deterministic interpretation that high-density hyperonic repulsion is important, but they also show that mass--radius information alone is insufficient for a precise determination of the hyperonic EDF parameters.

Several extensions remain for future work.
Mixed-phase treatments should also be examined separately; in
particular, a Gibbs construction with global charge neutrality may
modify the effective EOS in the coexistence region.
Mixed-phase treatments should also be examined separately; in particular, a Gibbs construction with global charge neutrality may modify the effective EOS in the coexistence region. Further work should also include simultaneous variations of the $\Lambda NN$ and $\Lambda\Lambda N$ sectors and combine neutron-star observables with hypernuclear and nuclear constraints in a unified inference framework.

\section*{Acknowledgments}

This work is supported by JSPS KAKENHI, Grant No.~JP23K03410, No.~JP23K25864,
and No.~JP25H01269.

\bibliography{ref}

\end{document}